\documentclass[a4paper,11pt]{article}

\usepackage{jcappub}
\usepackage[T1]{fontenc}
\usepackage[english]{babel}
\usepackage{float}
\usepackage{array}

\newcommand{\PreserveBackslash}[1]{\let\temp=\\#1\let\\=\temp}
\newcolumntype{C}[1]{>{\PreserveBackslash\centering}p{#1}}

\newcommand{\deltafct}[1]{\delta_\text{D}^{(#1)}}
\newcommand{\etain}{{\eta_\text{in}}}
\newcommand{\Pin}[1]{P_{#1}^\text{in}}
\newcommand{\gR}[1]{g_{#1}^R}
\newcommand{\gRtilde}[1]{\tilde{g}_{#1}^R}
\newcommand{\kfs}{k_\text{fs}}
\newcommand{\kfseq}{k_\text{fs,eq}}
\newcommand{\sigmahatin}{\hat{\sigma}_\text{in}}

\title{Evolution of dark matter velocity dispersion}

\author{Alaric Erschfeld}
\author{and Stefan Floerchinger}
\affiliation{Institut f\"{u}r Theoretische Physik, Ruprecht-Karls-Universit\"{a}t Heidelberg, \\ Philosophenweg 16, D-69120 Heidelberg, Germany}

\emailAdd{a.erschfeld@thphys.uni-heidelberg.de}
\emailAdd{stefan.floerchinger@thphys.uni-heidelberg.de}

\abstract{Cosmological perturbation theory for the late Universe dominated by dark matter is extended beyond the perfect pressureless fluid approximation by taking the dark matter velocity dispersion tensor as an additional field into account. A proper tensor decomposition of the latter leads to two additional scalar fields, as well as a vector and a tensor field. Most importantly, the trace of the velocity dispersion tensor can have a spatially homogeneous expectation value. While it decays at early times, we show that a back-reaction effect quadratic in perturbations makes it grow strongly at late times. We compare sterile neutrinos as a candidate for comparatively warm dark matter to weakly interacting massive particles as a rather cold dark matter candidate and show that the late time growth of velocity dispersion is stronger for the latter. Another feature of a non-vanishing velocity dispersion tensor is that it can account for multiple streams within the fluid description and thereby allows to treat times and scales beyond shell-crossing.}

\begin{document}

\maketitle
\flushbottom

\section{Introduction}
The matter distribution throughout the Universe shows structures on a wide range of scales. Surveys indicate that matter is distributed homogeneous and isotropic on the larges scales, commonly referred to as the cosmological principle. But observations also show a rich variety of dense cosmological objects, from stars to galaxies, galaxy groups to galaxy clusters, forming filaments separated by immense voids making up the so-called cosmic web. It is assumed that these structures are the outcome of gravitational instabilities of near to scale-invariant and Gaussian initial density fluctuations in the matter distribution in an early epoch where the Universe is dominated by non-relativistic matter, dubbed \emph{cold dark matter}. The collapse of matter is a competition between the expansion of the Universe and gravity, and current observations favour a scenario in which small scale structures collapse and virialise earlier than larger scales structures, also known as the bottom-up scenario \cite{SDSS_2004, 2dFGRS_2005}.
\par
From a theoretical point of view, structure formation can be described by following the evolution of an ensemble of self-gravitating dark matter particles. Commonly this is done by considering Newtonian gravity because the scales under consideration are much smaller than the Hubble horizon, where general relativistic corrections need to be taken into account. One often uses a description in terms of classical kinetic theory for the phase-space distribution of dark matter particles. This uses the collisionless limit of the Boltzmann equation with gravitational fields, generically known as Vlasov equation. The further description is done in either of two frames. Eulerian coordinates are fixed to the expanding Friedmann-Lema\^{i}tre-Robertson-Walker background cosmology, while Lagrangian coordinates follow the trajectories of individual dark matter particles or the so-called displacement field \cite{zeldovich_1970}. Working in the former, one can consider \textit{moments} or \textit{cumulants} of the dark matter velocity distribution to obtain an infinite tower of coupled equations. Often they are truncated by setting the velocity dispersion tensor and higher cumulants to zero. The resulting equations describe dark matter as a perfect pressureless fluid. This is the so-called \emph{single-stream approximation} since it assigns a unique velocity value to each point in space. Although the evolution equations are non-linear, these non-linearities are assumed to be small at early times and large scales and can therefore be treated perturbatively. For a review of standard cosmological perturbation theory we refer to reference \cite{bernardeau_2002}.
\par
Linear perturbation theory has had a tremendous success in describing the evolution of cosmic fields at early times and large scales. However, at later times and smaller scales, gravitational collapse becomes non-linear. A common way to deal with non-linearities in structure formation are numerical $N$-body simulations \cite{davis_1985, springel_2005}. While these allow to study structure formation in detail over a large amount of scales, they are limited by computational power. For this reason and to gain further insight into the mechanisms at play, different theoretical schemes have been put forward. Higher orders in perturbation theory have been calculated \cite{fry_1984, goroff_1986, makino_1992, jain_1994, scoccimarro_1996_1, scoccimarro_1996_2, takahashi_2008, shoji_2009, blas_2013_1}, various resummation schemes have been devised \cite{valageas_2004, crocce_2006_1, crocce_2006_2, mcdonald_2007, valageas_2007, matarrese_2007, izumi_2007, taruya_2008, crocce_2008, matsubara_2008_1, valageas_2008, bernardeau_2008_1, matsubara_2008_2, pietroni_2008, bernardeau_2008_2, taruya_2009, bernardeau_2010, anselmi_2011, bernardeau_2012_1, rampf_2012_1, rampf_2012_2, bernardeau_2012_2, taruya_2012, anselmi_2012, bernardeau_2013, blas_2016_1, blas_2016_2, ivanov_2018}, effective theories for a description of a larger range of scales have been developed \cite{enqvist_2011, baumann_2012, carrasco_2012, pajer_2013, mercolli_2014, porto_2014, carrasco_2014_1, carrasco_2014_2, senatore_2015_1, baldauf_2015_1, angulo_2015_1, angulo_2015_2, senatore_2015_2, assassi_2015, blas_2015, baldauf_2015_2, abolhasani_2016, bertolini_2016_1, bertolini_2016_2, floerchinger_2016, bertolini_2016_3, cataneo_2017, lewandowski_2017_1, munshi_2017, lewandowski_2017_2, bella_2017, senatore_2018_1, cusin_2018, bose_2018, senatore_2018_2} and the renormalisation group for a (statistical) field theoretic description has been studied \cite{matarrese_2007, matarrese_2008, floerchinger_2017}.
\par
However, to the extend that these methods are based on the single-stream approximation, they are ultimately limited to early times or very large scales. The approximation breaks down when multiple streams of matter coexist at the same region in space, a phenomenon known as \emph{shell-crossing}. By definition, the single-stream approximation can not account for multiple velocities at the same point in space. In Lagrangian coordinates, the gravitational force becomes intrinsically non-local after stream crossing and thus the density field can no longer be approximated as an expansion of the determinant of the local deformation tensor.
\par
There have been a number of works addressing this fundamental flaw in the description of dark matter. In an approach to estimate the effects of shell-crossing in Eulerian perturbation theory, reference \cite{pueblas_2009} computed the back-reaction of vorticity and velocity dispersion generation on the density and velocity-divergence fields using numerical simulations. In references \cite{pietroni_2012} and \cite{manzotti_2014}, velocity dispersion has been accounted for explicitly in the derivation from the Vlasov equation as source terms which are provided by numerical $N$-body simulations. Similarly, reference \cite{porto_2014} formulated an effective field theory by extending Lagrangian perturbation theory with sources. These encode small scale effects such as shell-crossing and the long wavelength expansion is fitted using numerical simulations. All these approaches are in some way `effective' in the sense that they encode the small scale effects in parameters which need to be fitted using numerical simulations. In order to extent the description of dark matter to times and scales beyond shell-crossing, a Vlasov-Einstein system \cite{piattella_2013, piattella_2016} as well as Schr\"{o}dinger-Poisson methods have been studied \cite{widrow_1993, coles_2003, uhlemann_2014, garny_2018, uhlemann_2018}. More recently, advances have been made in the Lagrangian picture of the Vlasov-Poisson system \cite{aviles_2016, cusin_2016}, especially in terms of 1+1 dimensional gravity \cite{mcquinn_2016, taruya_2017, rampf_2017, mcdonald_2018, pajer_2018, pietroni_2018}, where Zel'dovich dynamics is exact up to shell-crossing, and recently even beyond the 1+1 dimensional setting \cite{mcdonald_2018}. Rather than pursuing down this road we extent the Eulerian framework of the Vlasov-Poisson system, similarly as has been started in reference \cite{mcdonald_2011}.
\par
Let us recall that the insufficiency of the single-stream approximation is due to the truncation of the infinite Vlasov hierarchy. The single-stream approximation seems to be self-consistent if velocity dispersion is absent, but is unstable under perturbations. Including the second cumulant of the velocity distribution, namely velocity dispersion, introduces new degrees of freedom which allow to account for multiple streams within the fluid description. While velocity dispersion effects should be relatively small for cold dark matter, their corrections are expected to kick in at small enough scales. As we show, even a small seed of velocity dispersion can lead to accountable corrections at late times and small scales. 
\par
The other aspect we want to start investigating in this paper are cosmological vector and tensor fields. The single-stream approximation is described by two scalar fields, namely the density and velocity-divergence, while the only vector field present is the curl of the velocity field, the vorticity field. It is well known that at linear level the vorticity field decays due to the expansion of the Universe. In the single-stream approximation there are no non-linear terms which source vector fields from scalar initial conditions only. However, after including a non-vanishing velocity dispersion tensor we find that vector and tensor fields are non-linearly sourced by scalar fields. In the present paper we treat the full set of four scalar, two vector and one tensor matter degrees of freedom on linear level. In future follow-up work we plan to address non-linear dynamics in this setup.

\section{The Vlasov-Poisson system for dark matter}
\subsection{Dark matter distribution function, moments and cumulants}
We assume that dark matter consists of non-relativistic, classical point particles of mass $m$ which interact only gravitationally, at least to leading approximation. We work in an approximation where the metric is in the conformal Newtonian gauge \cite{mukhanov_1992}
\begin{equation}
ds^2 = a^2(\tau) \, \Big[ - \big( 1 + 2 \, \phi(\tau, \mathbf{x}) \big) \, d\tau^2 + \big( 1 - 2 \, \phi(\tau, \mathbf{x}) \big) \, d\mathbf{x}^2 \Big] \; ,
\label{eq:line_element}
\end{equation}
with the scale factor $a(\tau)$ and peculiar Newtonian gravitational potential $\phi(\tau, \mathbf{x})$. The conformal time $\tau$ is related to standard cosmic time $t$ by $dt = a(\tau) \, d\tau$. The gravitational potential is assumed to be relatively small on the scales we are interested in and for $\phi(\tau, \mathbf{x}) = 0$ the space-time has the symmetries of a flat Friedmann-Lema\^{i}tre-Robertson-Walker cosmology.\footnote{These symmetries actually become \emph{statistical symmetries} for the full theory where the expectation values are either taken as spatial averages over large enough volumes or for conveniently defined ensembles of cosmologies.} For a cosmology accounting for a cosmological constant and one dark matter species only, the evolution of the scale factor is determined by the Friedmann equations \cite{friedman_1922}
\begin{equation}
\frac{\dot{\mathcal{H}}(\tau)}{\mathcal{H}^2(\tau)} = 1 - \frac{3}{2} \, \Omega_m(\tau) \; , \qquad 1 = \Omega_m(\tau) + \Omega_\Lambda(\tau) \; ,
\label{eq:friedmann_equations}
\end{equation}
assuming dark matter to be pressureless at large enough scales. Here $\mathcal{H}(\tau) \equiv \dot{a}(\tau) / a(\tau)$ is the conformal Hubble parameter where we denote partial derivatives with respect to $\tau$ by a dot. Moreover, $\Omega_m (\tau)$ and $\Omega_\Lambda (\tau)$ are the dark matter and cosmological constant density parameters at time $\tau$, respectively. On scales much smaller than the Hubble horizon $1 / \mathcal{H}(\tau)$, the gravitational potential is in first order approximation subject to Poisson's equation
\begin{equation}
\Delta_\mathbf{x} \phi = \frac{3}{2} \, \mathcal{H}^2 \, \Omega_m \, \delta \; .
\label{eq:poissons_equation}
\end{equation}
To simplify notation we suppress time and space arguments when unambiguous. The dark matter mass density fluctuation field $\delta(\tau, \mathbf{x})$ is defined below.
\par
Dark matter is described in the framework of a kinetic theory approximation by the one-particle phase-space distribution function $f(\tau, \mathbf{x}, \mathbf{p})$ such that the number of particles in the comoving phase-space volume $d^3x \, d^3p$ at $(\mathbf{x}, \mathbf{p})$ and time $\tau$ is $f(\tau, \mathbf{x}, \mathbf{p}) \, d^3x \, d^3p$. Single dark matter particles move along the characteristics \cite{peebles_1980}
\begin{equation}
\frac{d\mathbf{x}}{d\tau} = \frac{\mathbf{p}}{a(\tau) \, m} \; , \qquad \frac{d\mathbf{p}}{d\tau} = - a(\tau) \, m \, \nabla_\mathbf{x} \phi(\tau, \mathbf{x}) \; ,
\end{equation}
such that the dark matter distribution function is conserved along them which leads to the Vlasov equation
\begin{equation}
\partial_\tau f + \frac{\mathbf{p}}{a \, m} \cdot \nabla_{\mathbf{x}} f - a \, m \, \nabla_{\mathbf{x}} \phi \cdot \nabla_{\mathbf{p}} f = 0 \; .
\label{eq:vlasov_equation}
\end{equation}
The Vlasov-Poisson system of equations \eqref{eq:poissons_equation} and \eqref{eq:vlasov_equation} is closed by the dark matter mass density field
\begin{equation}
\rho(\tau, \mathbf{x}) = \frac{m}{a^3(\tau)} \int_{\mathbb{R}^3} d^3p \, f(\tau, \mathbf{x}, \mathbf{p}) \; .
\label{eq:mass_density_field}
\end{equation}
In the following it is convenient to split the density field into a spatially homogeneous expectation value $\bar{\rho} (\tau) \equiv \left\langle \rho(\tau, \mathbf{x}) \right\rangle \propto a^{-3}(\tau)$ which determines the dark matter density parameter $\Omega_m (\tau)$ and the locally varying mass density fluctuation field
\begin{equation}
\delta(\tau, \mathbf{x}) \equiv \frac{\rho(\tau, \mathbf{x}) - \bar{\rho}(\tau)}{\bar{\rho}(\tau)} \; .
\label{eq:density_fluctuation_field}
\end{equation}
The latter enters Poisson's equation \eqref{eq:poissons_equation} such that the Vlasov-Poisson system of equations is non-linear.
\par
The Vlasov-Poisson system of equations specifies the time evolution of the distribution function $f(\tau, \mathbf{x}, \mathbf{p})$ and the metric parametrised by the gravitational potential $\phi(\tau, \mathbf{x})$. A theoretical drawback when considering the distribution function is that it is a function of seven variables which makes it naturally difficult to solve the Vlasov-Poisson system of equations. Further, in a statistical description where one considers expectation values $\left\langle f(\tau, \mathbf{x}, \mathbf{p}) \right\rangle$ or correlation functions $\left\langle f(\tau, \mathbf{x}, \mathbf{p}) \, f(\tau^\prime, \mathbf{x}^\prime, \mathbf{p}^\prime) \right\rangle$, the symmetries reduce the complexity only little. On the other side, from an observational point of view, one is often not interested in the full phase-space distribution function but rather in \emph{moments} and \emph{cumulants} with respect to the momentum argument. The zeroth moment of the distribution function is the mass density field \eqref{eq:mass_density_field}. The first moment is given by the momentum density field
\begin{equation}
\rho(\tau, \mathbf{x}) \, u_i(\tau, \mathbf{x}) = \frac{m}{a^3(\tau)} \int_{\mathbb{R}^3} d^3p \, \frac{p_i}{a(\tau) \, m} \, f(\tau, \mathbf{x}, \mathbf{p}) \; ,
\label{eq:peculiar_velocity_field}
\end{equation}
where $u_i(\tau, \mathbf{x})$ is the peculiar velocity field. The second moment is the momentum flux density field
\begin{equation}
\rho(\tau, \mathbf{x}) \, u_i(\tau, \mathbf{x}) \, u_j(\tau, \mathbf{x}) + T_{ij}(\tau, \mathbf{x}) = \frac{m}{a^3(\tau)} \int_{\mathbb{R}^3} d^3p \, \frac{p_i p_j}{a^2(\tau) \, m^2} \, f(\tau, \mathbf{x}, \mathbf{p}) \; ,
\label{eq:velocity_dispersion_tensor_field}
\end{equation}
where $T_{ij}(\tau, \mathbf{x})$ is the stress tensor field. Higher moments of the distribution function can be defined very similar to \eqref{eq:mass_density_field}, \eqref{eq:peculiar_velocity_field} and \eqref{eq:velocity_dispersion_tensor_field} and in close analogy to the usual definitions for probability distributions. It is convenient to split the stress tensor field $T_{ij}(\tau, \mathbf{x}) = \rho(\tau, \mathbf{x}) \, \sigma_{ij}(\tau, \mathbf{x})$ where $\sigma_{ij}(\tau, \mathbf{x})$ is the peculiar velocity dispersion tensor field which quantifies the deviation of particle velocities from the velocity field $u_i(\tau, \mathbf{x})$. Since only the density fluctuation field $\delta(\tau, \mathbf{x})$ enters Poisson's equation \eqref{eq:poissons_equation}, it is convenient to discuss moments and cumulants with respect to the momentum argument of the redefined distribution function
\begin{equation}
\tilde{f}(\tau, \mathbf{x}, \mathbf{p}) \equiv \frac{1}{\bar{\rho}(\tau)} \, \frac{m}{a^3(\tau)} \, f(\tau, \mathbf{x}, \mathbf{p}) \; ,
\end{equation}
which obeys the Vlasov equation \eqref{eq:vlasov_equation} because it is linear in $f(\tau, \mathbf{x}, \mathbf{p})$ and the density background decays $\bar{\rho}(\tau) \propto a^{-3}(\tau)$. Moments $m_{i_1 ... i_n}^{(n)}(\tau, \mathbf{x})$ of any order $n \in \mathbb{N}_0$ of the distribution function with respect to the momentum argument parametrise spatial matter degrees of freedom of the fluid. They can be obtained from the moment-generating function
\begin{equation}
M(\tau, \mathbf{x}; \mathbf{l}) \equiv \int_{\mathbb{R}^3} d^3p \, \exp \left\{ \frac{\mathbf{l} \cdot \mathbf{p}}{a(\tau) \, m} \right\} \tilde{f}(\tau, \mathbf{x}, \mathbf{p}) \; ,
\end{equation}
by applying derivatives with respect to $\mathbf{l}$ and evaluating at $\mathbf{l} = \mathbf{0}$,
\begin{equation}
m_{i_1 ... i_n}^{(n)}(\tau, \mathbf{x}) = \frac{\partial^n M(\tau, \mathbf{x}; \mathbf{l})}{\partial l_{i_1} ... \partial l_{i_n}} \bigg\vert_{\mathbf{l} = \mathbf{0}} \; ,
\label{eq:moments}
\end{equation}
where the indices $i_j \in \{ 1, 2, 3 \}$ are spatial and totally symmetric for all $j \in \{ 1, ..., n \}$. Similarly, cumulants $c_{i_1 ... i_n}^{(n)}(\tau, \mathbf{x})$ of the distribution function are statistically independent quantities which represent the connected part of the moments. They can be obtained from the cumulant-generating function $C(\tau, \mathbf{x}; \mathbf{l}) \equiv \ln \big( M(\tau, \mathbf{x}; \mathbf{l}) \big)$ in the same manner moments are derived from $M(\tau, \mathbf{x}; \mathbf{l})$ in equation \eqref{eq:moments}. The first few cumulants are given by
\begin{equation}
c^{(0)}(\tau, \mathbf{x}) = \ln \big( 1 + \delta(\tau, \mathbf{x}) \big) \; , \qquad c_i^{(1)}(\tau, \mathbf{x}) = u_i(\tau, \mathbf{x}) \; , \qquad c_{ij}^{(2)}(\tau, \mathbf{x}) = \sigma_{ij}(\tau, \mathbf{x}) \; .
\label{eq:cumulants}
\end{equation}
With the definition of the cumulant-generating function and the Vlasov equation \eqref{eq:vlasov_equation} one can derive the equation of motion for the cumulant-generating function \cite{pueblas_2009, uhlemann_2018},
\begin{equation}
\partial_\tau C + \mathcal{H} \, \mathbf{l} \cdot \nabla_\mathbf{l} C + \nabla_\mathbf{x} \cdot \nabla_\mathbf{l} C + \nabla_\mathbf{x} C \cdot \nabla_\mathbf{l} C + \mathbf{l} \cdot \nabla_\mathbf{x} \phi = 0 \; .
\label{eq:cumulant_vlasov_equation}
\end{equation}
From the structure of equation \eqref{eq:cumulant_vlasov_equation} one can already infer the most important properties of the time evolution of cumulants of any order $n$:
\begin{itemize}
\item The second term on the left-hand side generates the Hubble drag term accounting for the expansion of the Universe, which scales with the number of spatial indices $n$.
\item The third term on the left-hand side involves the next higher order cumulant, creating an infinity hierarchy of coupled equations.
\item The fourth term on the left-hand side is quadratic in the cumulants and generates a sum over the next higher and all lower order cumulants.
\item The last term on the left-hand side only contributes at $n = 1$ and couples Poisson's equation \eqref{eq:poissons_equation} to the evolution of the first cumulant.
\end{itemize}
To have a finite set of closed equations one needs to truncate the Vlasov hierarchy in an appropriate way. From the non-linear term in the equations of motion \eqref{eq:cumulant_vlasov_equation} it is clear, that the only (apparent) consistent truncation is neglecting all cumulants of order $n \ge 2$, because there are no source terms in the corresponding equations of motion which solely depend on lower order cumulants. This corresponds to assuming dark matter to behave like a perfect pressureless fluid and amounts to the generating functions
\begin{equation}
C(\tau, \mathbf{x}; \mathbf{l}) = c^{(0)}(\tau, \mathbf{x}) + l_i \, c_i^{(1)}(\tau, \mathbf{x}) \; , \qquad M(\tau, \mathbf{x}; \mathbf{l}) = \big[ 1 + \delta(\tau, \mathbf{x}) \big] \, e^{ \mathbf{l} \cdot \mathbf{u}(\tau, \mathbf{x}) } \; ,
\end{equation}
which correspond to the distribution function
\begin{equation}
\tilde{f}(\tau, \mathbf{x}, \mathbf{p}) = \big[ 1 + \delta(\tau, \mathbf{x}) \big] \, \deltafct{3} \big( \mathbf{p} - a(\tau) \, m \, \mathbf{u}(\tau, \mathbf{x}) \big) \; ,
\label{eq:single-stream_approximation_distribution_function}
\end{equation}
where $\delta_\text{D}^{(n)}(x)$ is the Dirac delta function in $n$ dimensions and we enforce the Einstein summation convention. This so-called \emph{single-stream approximation} assigns a single velocity field $\mathbf{u}(\tau, \mathbf{x})$ to each point in configuration space. While the single-stream approximation is self-consistent, it is unstable under perturbations. Any deviation from the distribution function \eqref{eq:single-stream_approximation_distribution_function} naturally generates all higher order moments \cite{pueblas_2009}. Further, the single-stream approximation can not account for multiple streams within the fluid and is therefore unable to describe the phenomenon of \emph{shell-crossing}, when trajectories of particles meet in configuration space. At this point the velocity field in equation \eqref{eq:single-stream_approximation_distribution_function} is multivalued and all roots of the Dirac delta function contribute. This amounts to a distribution function of the form
\begin{equation}
\tilde{f}(\tau, \mathbf{x}, \mathbf{p}) = \sum_{s} \big[ 1 + \delta_s(\tau, \mathbf{x}) \big] \, \deltafct{3} \big( \mathbf{p} - a(\tau) \, m \, \mathbf{u}_s(\tau, \mathbf{x}) \big) \; ,
\end{equation}
where $s$ runs over all streams with density and velocity field, $\delta_s(\tau, \mathbf{x})$ and $\mathbf{u}_s(\tau, \mathbf{x})$, respectively. From the corresponding cumulant-generating function it is clear that this form of distribution function induces cumulants of any order \cite{pueblas_2009}.
\par
In order to overcome the perfect pressureless fluid approximation we truncate the Vlasov hierarchy at the next higher cumulant, thereby accounting for velocity dispersion within the fluid. This amounts to the cumulant-generating function
\begin{equation}
C(\tau, \mathbf{x}; \mathbf{l}) = c^{(0)}(\tau, \mathbf{x}) + l_i \, c_i^{(1)}(\tau, \mathbf{x}) + \frac{1}{2} \, l_i \, l_j \, c_{ij}^{(2)}(\tau, \mathbf{x}) \; ,
\end{equation}
and thus to the moment-generating function
\begin{equation}
M(\tau, \mathbf{x}; \mathbf{l}) = \big[ 1 + \delta(\tau, \mathbf{x}) \big] \, e^{ l_i \, u_i(\tau, \mathbf{x}) + \frac{1}{2} \, l_i \, l_j \, \sigma_{ij}(\tau, \mathbf{x}) } \; .
\end{equation}
The corresponding distribution function is the three-dimensional normal distribution
\begin{equation}
\tilde{f}(\tau, \mathbf{x}, \mathbf{p}) = \frac{1 + \delta}{\sqrt{(2 \pi)^3 \, \det(a^2 \, m^2 \, \boldsymbol{\sigma})}} \, \exp \left\{- \frac{1}{2} \, \big( p_i - a \, m \, u_i \big) \, \frac{\sigma_{ij}^{-1}}{a^2 \, m^2} \, \big( p_j- a \, m \, u_j \big) \right\} \; ,
\label{eq:velocity_dispersion_distribution_function}
\end{equation}
where this form only holds if the isotropic velocity dispersion is sufficiently larger than anisotropic velocity dispersion, such that $\sigma_{ij}(\tau, \mathbf{x})$ is positive definite. A few comments are in order:
\begin{itemize}
\item The momenta of single dark matter particles are normal distributed around the mean velocity field $u_i(\tau, \mathbf{x})$ with covariance $\sigma_{ij}(\tau, \mathbf{x})$.
\item The distribution function \eqref{eq:velocity_dispersion_distribution_function} can capture the phenomenon of shell-crossing.
\item The single-stream approximation \eqref{eq:single-stream_approximation_distribution_function} is recovered in the limit $\sigma_{ij}(\tau, \mathbf{x}) \to 0$.
\item In thermal equilibrium we expect $u_i(\tau, \mathbf{x}) = 0$ and $\sigma_{ij}(\tau, \mathbf{x}) = \delta_{ij} \, k_\text{B} T / m$, such that momenta are Maxwell-Boltzmann distributed
\begin{equation}
\tilde{f}(\tau, \mathbf{x}, \mathbf{p}) = \; \frac{1 + \delta}{a^3 \, m^3} \left( \frac{m}{2 \pi k_\text{B} T} \right)^\frac{3}{2} \exp \left\{- \frac{m}{2 k_\text{B} T} \, \frac{\mathbf{p}^2}{a^2 \, m^2} \right\} \; .
\end{equation}
There are indications that a Maxwell-Boltzmann distribution also describes the dark matter velocity distribution within bond structures such as halos \cite{bahcall_1994_1,bahcall_1994_2}.
\end{itemize}
From the discussion of the equation of motion \eqref{eq:cumulant_vlasov_equation} we know that the Vlasov equation \eqref{eq:vlasov_equation} does not conserve this approximation, because higher order cumulants are dynamically generated. However, \emph{cum grano salis} we assume this truncation to be appropriate for times and scales where the dark matter distribution function is perturbed around thermal equilibrium sufficiently close to \eqref{eq:velocity_dispersion_distribution_function} and henceforth set all higher order cumulants zero.
\par
Thus, the set of equations we are taking into account are the continuity equation
\begin{equation}
\dot{\delta} + u_i \, \delta_{,i} + (1 + \delta) \, u_{i,i} = 0 \; ,
\label{eq:continuity_eq_v1}
\end{equation}
the Cauchy momentum equation
\begin{equation}
\dot{u}_i + \mathcal{H} \, u_i + u_j \, u_{i,j} + \sigma_{ij,j} + \sigma_{ij} \, \ln ( 1 + \delta )_{,j} + \phi_{,i} = 0 \; ,
\label{eq:cauchy_momentum_eq_v1}
\end{equation}
and the velocity dispersion equation
\begin{equation}
\dot{\sigma}_{ij} + 2 \mathcal{H} \, \sigma_{ij} + u_k \, \sigma_{ij,k} + \sigma_{jk} \, u_{i,k} + \sigma_{ik} \, u_{j,k} = 0 \; .
\label{eq:velocity_dispersion_eq_v1}
\end{equation}
Here we use a notation where a comma indicates partial derivates with respect to the components of $\mathbf{x}$.

\subsection{Statistical description and background-fluctuation splitting}
We are interested in a statistical description of dark matter formulated in terms of expectation values and correlation functions of the cosmological fields. The latter are given by the gravitational potential $\phi(\tau, \mathbf{x})$ as well as the matter fields defined by the zeroth moment \eqref{eq:mass_density_field} and the first two cumulants  \eqref{eq:cumulants} of the dark matter distribution function $f(\tau, \mathbf{x}, \mathbf{p})$.
\par
The statistical field theory we employ can be seen as describing an ensemble of cosmological histories with stochastic initial conditions or, equivalently for this purpose, an ensemble of different spatial subvolumes of a single cosmological history. We assume that the symmetries of the idealised flat Friedmann-Lema\^{i}tre-Robertson-Walker cosmology are realised as statistical symmetries for this field theory which concerns in particular spatial translations and rotations.
\par
In the following it is convenient to split the cosmological fields into a background and a fluctuation part. The background is taken to be spatially homogeneous and isotropic and corresponds to the expectation value of the fields. For the density field we have already done such a splitting in definition \eqref{eq:density_fluctuation_field}. For the velocity field $u_i(\tau, \mathbf{x})$ the statistical rotation symmetry forbids an expectation value such that it is a pure fluctuation field. For the velocity dispersion tensor field defined in \eqref{eq:velocity_dispersion_tensor_field} one may have a non-vanishing isotropic expectation value,
\begin{equation}
\left\langle \sigma_{ij}(\tau, \mathbf{x}) \right\rangle = \delta_{ij} \, \bar{\sigma}(\tau) \; .
\label{eq:background_velocity_dispersion}
\end{equation}
Note that $\bar{\sigma}(\tau)$ is positive semi-definite and in the following we assume $\bar{\sigma}(\tau) > 0$, although the value could be very small. Similar to the density fluctuation field \eqref{eq:density_fluctuation_field} we define the peculiar velocity dispersion tensor fluctuation field
\begin{equation}
\varsigma_{ij}(\tau, \mathbf{x}) \equiv \sigma_{ij}(\tau, \mathbf{x}) - \delta_{ij} \, \bar{\sigma}(\tau) \; .
\label{eq:peculiar_velocity_dispersion_tensor_field}
\end{equation}
The velocity dispersion background $\bar{\sigma}(\tau)$ obeys the expectation value of the trace of the velocity dispersion equation \eqref{eq:velocity_dispersion_eq_v1}, given by
\begin{equation}
\dot{\bar{\sigma}}(\tau) + 2 \mathcal{H}(\tau) \, \bar{\sigma}(\tau) + \frac{1}{3} \, \langle u_i(\tau, \mathbf{x}) \, \varsigma_{jj,i}(\tau, \mathbf{x}) \rangle + \frac{2}{3} \, \langle \varsigma_{ij}(\tau, \mathbf{x}) \, u_{i,j}(\tau, \mathbf{x}) \rangle = 0 \; .
\label{eq:background_velocity_dispersion_eq_v1}
\end{equation}
The evolution of the velocity dispersion background depends on equal-time two-point correlation functions of the velocity and the velocity dispersion fluctuation fields.
\par
Note that the expectation value \eqref{eq:background_velocity_dispersion} differs from a possible expectation value of the pressure of the dark matter fluid which would instead be given by
\begin{equation}
\left\langle T_{ij}(\tau, \mathbf{x}) \right\rangle = \delta_{ij} \, \bar{P} (\tau) \; .
\end{equation}
While one might wonder if directly working with the pressure background is more sensible, we do not see an a priori reason why this should be the case. In terms of the background-fluctuation splitting \eqref{eq:density_fluctuation_field} and \eqref{eq:peculiar_velocity_dispersion_tensor_field} the dark matter pressure and velocity dispersion background can be related
\begin{equation}
\bar{P} (\tau) = \bar{\rho}(\tau) \, \bar{\sigma}(\tau) + \frac{\bar{\rho}(\tau)}{3} \left\langle \delta(\tau, \mathbf{x}) \, \varsigma_{ii}(\tau, \mathbf{x}) \right\rangle \; .
\end{equation}
This pressure background enters the Friedmann equations \eqref{eq:friedmann_equations}, but is suppressed by powers of the speed of light.
\par
In terms of the background-fluctuation splitting \eqref{eq:background_velocity_dispersion} and \eqref{eq:peculiar_velocity_dispersion_tensor_field}, the Cauchy momentum equation \eqref{eq:cauchy_momentum_eq_v1} and velocity dispersion equation \eqref{eq:velocity_dispersion_eq_v1} can be written as
\begin{equation}
\dot{u}_i + \mathcal{H} u_i + u_j u_{i,j} + \bar{\sigma} \, \varsigma_{ij,j} + \bar{\sigma} \, \ln( 1 + \delta )_{,i} + \bar{\sigma} \, \varsigma_{ij} \ln(1 + \delta)_{,j} + \phi_{,i} = 0 \; ,
\label{eq:cauchy_momentum_eq_v2}
\end{equation}
and
\begin{equation}
\dot{\varsigma}_{ij} + 2 \mathcal{H} \varsigma_{ij} + u_k \varsigma_{ij,k} + \bar{\sigma} (u_{i,j} + u_{j,i}) + \varsigma_{jk} u_{i,k} + \varsigma_{ik} u_{j,k} + \dot{\bar{\sigma}} + 2 \mathcal{H} \bar{\sigma} = 0 \; .
\label{eq:velocity_dispersion_eq_v2}
\end{equation}
The last term on the left side of equation \eqref{eq:velocity_dispersion_eq_v2} is a background term which, together with the evolution equation \eqref{eq:background_velocity_dispersion_eq_v1}, ensures that $\langle \dot{\varsigma}_{ij}(\tau, \mathbf{x}) \rangle = 0$. Note that $\bar{\sigma}(\tau)$ enters the equations of motion \eqref{eq:cauchy_momentum_eq_v2} and \eqref{eq:velocity_dispersion_eq_v2} and that the evolution equation \eqref{eq:background_velocity_dispersion_eq_v1} depends on the two-point correlation function of the fluctuation fields. In this sense, the background and fluctuation fields are coupled and need to be solved simultaneously.

\subsection{Scalar, vector and tensor decomposition}
For further analysis we decompose the matter fields according to their transformation properties under spatial rotations into scalar, vector and tensor fields. This is most conveniently done by moving to Fourier space using the convention
\begin{equation}
f (\mathbf{k}) \equiv \int_{\mathbf{x}} \; e^{- i \mathbf{k} \cdot \mathbf{x}} \, f(\mathbf{x}) \; , \qquad f(\mathbf{x}) \equiv \int_{\mathbf{k}} e^{i \mathbf{k} \cdot \mathbf{x}} \, f(\mathbf{k}) \; ,
\end{equation}
for any integrable field $f(\mathbf{x})$ where we abbreviate configuration and Fourier space integrals by
\begin{equation}
\int_{\mathbf{x}} \equiv \int_{\mathbb{R}^3} d^3x \; , \qquad \int_{\mathbf{k}} \equiv \int_{\mathbb{R}^3} \frac{d^3k}{(2 \pi)^3} \; .
\end{equation}
The velocity fluctuation field $u_i(\tau, \mathbf{k})$ can be decomposed into an irrotational and a solenoidal vector field. The irrotational part of the decomposition can be parametrised in terms of the scalar peculiar velocity-divergence field
\begin{equation}
\theta(\tau, \mathbf{k}) \equiv i k_j \, u_j(\tau, \mathbf{k}) \; ,
\end{equation}
and describes the potential flow of the dark matter fluid. The remaining solenoidal part can be parametrised by the peculiar vorticity field
\begin{equation}
\omega_j(\tau, \mathbf{k}) \equiv \varepsilon_{jkl} \, i k_k \, u_l (\tau, \mathbf{k}) \; ,
\end{equation}
which quantifies the local rotation of fluid elements. The vorticity field $\omega_i(\tau, \mathbf{k})$ is a pseudovector field which is transverse to the flow direction $\mathbf{k}$. In terms of the velocity-divergence and vorticity fields, the decomposition takes the form
\begin{equation}
u_j(\tau, \mathbf{k}) = - \frac{i k_j}{k^2} \, \theta(\tau, \mathbf{k}) + \varepsilon_{jkl} \, \frac{i k_k}{k^2} \, \omega_l(\tau, \mathbf{k}) \; .
\label{eq:decomposed_peculiar_velocity_field}
\end{equation}
Similarly, the symmetric second-rank velocity dispersion tensor fluctuation field $\varsigma_{ij}(\tau, \mathbf{k})$ can be decomposed into two scalar fields, a solenoidal vector field and a symmetric, transverse and traceless second-rank tensor field \cite{lifshitz_1946}. We parametrise the scalar fields in terms of the trace field
\begin{equation}
\varsigma(\tau, \mathbf{k}) \equiv \frac{\delta_{ij}}{3} \, \varsigma_{ij} ( \tau, \mathbf{k} ) \; ,
\label{eq:scalar_isotropic_velocity_dispersion_fluctuation_field}
\end{equation}
and the off-trace field
\begin{equation}
\vartheta(\tau, \mathbf{k}) \equiv \left ( \frac{k_i k_j}{k^2} - \frac{\delta_{ij}}{3} \right ) \varsigma_{ij} ( \tau, \mathbf{k} ) \; ,
\label{eq:scalar_anisotropic_velocity_dispersion_fluctuation_field}
\end{equation}
which quantify isotropic and anisotropic velocity dispersion, respectively. The solenoidal vector field we parametrise by
\begin{equation}
\vartheta_i(\tau, \mathbf{k}) \equiv \varepsilon_{ijk} \, \frac{k_j k_l}{k^2} \, \varsigma_{kl} ( \tau, \mathbf{k} ) \; ,
\label{eq:vector_velocity_dispersion_field}
\end{equation}
and the symmetric, transverse and traceless second-rank tensor field by
\begin{equation}
\vartheta_{ij} ( \tau, \mathbf{k} ) \equiv \Delta_{ijkl}(\mathbf{k}) \, \varsigma_{kl} ( \tau, \mathbf{k} ) \; ,
\label{eq:tensor_velocity_dispersion_field}
\end{equation}
both of which describe anisotropic velocity dispersion degrees of freedom. Here we introduced the symmetric, transverse and traceless projector
\begin{equation}
\Delta_{ijkl} (\mathbf{k}) \equiv \frac{1}{2} \, \Big[ \Delta_{ik}(\mathbf{k}) \, \Delta_{jl}(\mathbf{k}) + \Delta_{il}(\mathbf{k}) \, \Delta_{jk}(\mathbf{k}) - \Delta_{ij}(\mathbf{k}) \, \Delta_{kl}(\mathbf{k}) \Big] \; ,
\end{equation}
defined in terms of the transverse projector
\begin{equation}
\Delta_{ij} (\mathbf{k}) \equiv \delta_{ij} - \frac{k_i k_j}{k^2} \; .
\end{equation}
The decomposition of the velocity dispersion tensor fluctuation field $\varsigma_{ij}(\tau, \mathbf{k})$ can then be written as
\begin{equation}
\begin{aligned}
\varsigma_{ij}(\tau, \mathbf{k}) = \delta_{ij} \, \varsigma(\tau, \mathbf{k}) + \frac{3}{2} \left( \frac{k_i k_j}{k^2} - \frac{\delta_{ij}}{3} \right) \vartheta(\tau, \mathbf{k}) &- \frac{(\varepsilon_{ikl} \, k_j + \varepsilon_{jkl} \, k_i) \, k_k}{k^2} \, \vartheta_l(\tau, \mathbf{k}) \\
&+ \Delta_{ijkl}(\mathbf{k}) \, \vartheta_{kl}(\tau, \mathbf{k}) \; .
\end{aligned}
\label{eq:decomposed_peculiar_velocity_dispersion_tensor_field}
\end{equation}
All ten matter degrees of freedom are described by the four scalar fields $\big( \delta, \theta, \varsigma, \vartheta \big)$, the two solenoidal vector fields $\big( \omega_i, \vartheta_i \big)$ and the symmetric, transverse and traceless second-rank tensor field $\vartheta_{ij}$.
\par
In order to rewrite the evolution equation \eqref{eq:background_velocity_dispersion_eq_v1} in terms of expectation values of the decomposed fluctuation fields it is convenient to define the equal-time power spectra
\begin{equation}
\begin{aligned}
(2 \pi)^3 \, \delta_\text{D}^{(3)}(\mathbf{k} + \mathbf{k}^\prime) \, P_{\varsigma \theta}(\tau, k) &\equiv \left\langle \varsigma(\tau, \mathbf{k}) \; \theta(\tau, \mathbf{k}^\prime) \right\rangle \; , \\
(2 \pi)^3 \, \delta_\text{D}^{(3)}(\mathbf{k} + \mathbf{k}^\prime) \, P_{\vartheta \theta}(\tau, k) &\equiv \left\langle \vartheta(\tau, \mathbf{k}) \; \theta(\tau, \mathbf{k}^\prime) \right\rangle \; , \\
(2 \pi)^3 \, \delta_\text{D}^{(3)}(\mathbf{k} + \mathbf{k}^\prime) \, \Delta_{ij}(\mathbf{k}) \, P_{\vartheta \omega}(\tau, k) &\equiv \left\langle \vartheta_i(\tau, \mathbf{k}) \; \omega_j(\tau, \mathbf{k}^\prime) \right\rangle \; .
\end{aligned}
\end{equation}
The evolution equation \eqref{eq:background_velocity_dispersion_eq_v1} can then be rewritten to
\begin{equation}
\dot{\bar{\sigma}}(\tau) + 2 \mathcal{H}(\tau) \, \bar{\sigma}(\tau) - Q(\tau) = 0 \; ,
\label{eq:background_velocity_dispersion_eq_v2}
\end{equation}
where we defined the dimensionless function
\begin{equation}
Q(\tau) \equiv \frac{1}{3} \int_\mathbf{q} \Big[ P_{\varsigma \theta}(\tau, q) - 2 \, P_{\vartheta \theta}(\tau, q) - 4 \, P_{\vartheta \omega}(\tau, q) \Big] \; ,
\label{eq:definition_Q}
\end{equation}
From equation \eqref{eq:background_velocity_dispersion_eq_v2} it is evident that the velocity dispersion background is sourced by the covariance of the velocity and velocity dispersion fields, or, physically speaking, by inhomogeneities in the velocity and velocity dispersion distribution. In this sense we henceforth refer to the function $Q(\tau)$ as source function.
\par
To cast the Cauchy momentum equation \eqref{eq:cauchy_momentum_eq_v2} into a form which is a polynomial of finite degree in the fluctuation fields, we expand the logarithmic density field $\ln( 1+\delta(\tau, \mathbf{x}))$ around the background configuration $\delta(\tau, \mathbf{x}) = 0$ by a Taylor series up to terms cubic in $\delta(\tau, \mathbf{x})$ and neglect all higher orders.\footnote{Higher order terms in the fluctuation fields do not enter one-loop calculations within the framework of renormalised cosmological perturbation theory \cite{crocce_2006_1}.} By splitting the equations of motion \eqref{eq:continuity_eq_v1}, \eqref{eq:cauchy_momentum_eq_v2} and \eqref{eq:velocity_dispersion_eq_v2} for $\mathbf{k} \ne 0$ according to the decompositions \eqref{eq:decomposed_peculiar_velocity_field} and \eqref{eq:decomposed_peculiar_velocity_dispersion_tensor_field} and using Poisson's equation \eqref{eq:poissons_equation} as well as the  evolution equation \eqref{eq:background_velocity_dispersion_eq_v2} one is left with the continuity equation
\begin{equation}
\dot{\delta} + \theta + I_\delta = 0 \; ,
\label{eq:continuity_eq_v2}
\end{equation}
the velocity-divergence equation
\begin{equation}
\dot{\theta} + \mathcal{H} \theta - k^2 \varsigma - k^2 \vartheta  + \left( \frac{3}{2} \, \mathcal{H}^2 \Omega_m - k^2 \bar{\sigma} \right) \delta + I_\theta + J_\theta = 0 \; ,
\label{eq:velocity-divergence_equation}
\end{equation}
the vorticity equation
\begin{equation}
\dot{\omega}_i + \mathcal{H} \omega_i - k^2 \vartheta_i + I_\omega^i + J_\omega^i = 0 \; ,
\label{eq:vorticity_equation}
\end{equation}
and the four velocity dispersion equations
\begin{equation}
\dot{\varsigma} + 2 \mathcal{H} \varsigma + \frac{2}{3} \, \bar{\sigma} \, \theta + I_\varsigma = 0 \; ,
\label{eq:trace_velocity_dispersion_eq}
\end{equation}
\begin{equation}
\dot{\vartheta} + 2 \mathcal{H} \vartheta + \frac{4}{3} \, \bar{\sigma} \, \theta + I_{\vartheta} = 0 \; ,
\end{equation}
\begin{equation}
\dot{\vartheta}_i + 2 \mathcal{H} \vartheta_i + \bar{\sigma} \, \omega_i + I_{\vartheta}^i = 0 \; ,
\end{equation}
\begin{equation}
\dot{\vartheta}_{ij} + 2 \mathcal{H} \vartheta_{ij} + I_{\vartheta}^{ij} = 0 \; .
\label{eq:tensor_velocity_dispersion_eq}
\end{equation}
Here we abbreviated terms that are quadratic and cubic in the fluctuation fields which are listed in appendix \ref{non-linear_terms_of_the_equations_of_motion}.
\par
The assumption of a non-vanishing $\bar{\sigma}(\tau)$ breaks the apparent self-consistency of the single-stream approximation since already at linear level the velocity dispersion fields $\varsigma(\tau, \mathbf{x})$, $\vartheta(\tau, \mathbf{x})$ and $\vartheta_i(\tau, \mathbf{x})$ are sourced by the velocity fields $\theta(\tau, \mathbf{x})$ and $\omega_i(\tau, \mathbf{x})$, respectively. Further, the single-stream approximation does not allow to generate the vorticity field from scalar initial conditions because $C_{\theta \omega}^{ij}(\mathbf{k}_1, \mathbf{k}_2)$ and $C_{\omega \omega}^{ijk}(\mathbf{k}_1, \mathbf{k}_2)$ are the only vertices present. In contrast, by including velocity dispersion, it is possible to source vector and tensor fields non-linearly from scalar initial conditions through the vertices $C_{\varsigma \delta}^i(\mathbf{k}_1, \mathbf{k}_2)$, $C_{\vartheta \delta}^i(\mathbf{k}_1, \mathbf{k}_2)$, $C_{\varsigma \delta \delta}^i(\mathbf{k}_1, \mathbf{k}_2, \mathbf{k}_3)$, $C_{\vartheta \delta \delta}^i(\mathbf{k}_1, \mathbf{k}_2, \mathbf{k}_3)$, $F_{\varsigma \theta}^i(\mathbf{k}_1, \mathbf{k}_2)$, $F_{\vartheta \theta}^i(\mathbf{k}_1, \mathbf{k}_2)$, $G_{\varsigma \theta}^{ij}(\mathbf{k}_1, \mathbf{k}_2)$, $G_{\vartheta \theta}^{ij}(\mathbf{k}_1, \mathbf{k}_2)$.
\par
Note that $\bar{\sigma}(\tau)$ enters the velocity-divergence and vorticity equations \eqref{eq:velocity-divergence_equation} and \eqref{eq:vorticity_equation} in the combination $k^2 \bar{\sigma}(\tau)$, as is required by dimensional arguments. This introduces a scale dependence even at linear order in the fluctuation fields which is crucial for the convergence of the source function $Q(\tau)$ as we explain in section \ref{sec:results}.

\subsection{Compact notation}
\label{sec:compact_notation}
To cast the equations of motion \eqref{eq:continuity_eq_v2} -- \eqref{eq:tensor_velocity_dispersion_eq} into a more symmetric and compact form we introduce the time evolution parameter
\begin{equation}
\eta(\tau) \equiv \ln \left( \frac{D_+(\tau)}{D_+(\tau_\text{in})} \right) \; ,
\label{eq:time_evolution_parameter}
\end{equation}
corresponding to the number of $e$-folds of the linear growth function in the single-stream approximation \cite{silveira_1994}
\begin{equation}
D_+(\tau) = a(\tau) \, {}_2 F_1 \left ( \frac{1}{3} \, , \; 1 \, ; \; \frac{11}{6} \, ; \; - a^3 (\tau) \, \frac{1 - \Omega_{m,0}}{\Omega_{m,0}} \right ) \; ,
\label{eq:linear_growth_function_single-stream_approximation}
\end{equation}
normalised to unity at some initial time $\tau_\text{in}$.\footnote{We assume a cosmology with a cosmological constant and one sufficiently cold dark matter species without radiative component.} Here the growth function is given in terms of the Gaussian hypergeometric function ${}_2 F_1(a, b; c; x)$ and the dark matter density parameter is written as $\Omega_m(\tau) = \Omega_{m,0} / a^3(\tau)$ where $\Omega_{m,0}$ is the value at $a(0) = 1$, corresponding to today. Further it is useful to define the dimensionless function
\begin{equation}
f(\tau) \equiv \frac{\partial \ln (D_+(\tau))}{\partial \ln (a(\tau))} \; ,
\end{equation}
which parametrises the logarithmic deviation of $D_+(\tau)$ from the Einstein-de Sitter linear growth function $a(\tau)$ and obeys the evolution equation \cite{pietroni_2012}
\begin{equation}
\partial_{\eta} \ln(f \, \mathcal{H}) = \frac{3}{2} \frac{\Omega_m}{f^2} - \frac{1}{f} - 1 \; .
\end{equation}
In the following we use $\eta(\tau)$ rather than $\tau$ which are related through definition \eqref{eq:time_evolution_parameter}.
\par
Grouping scalar, vector and tensor fields in a condensed notation, we define the field vector
\begin{equation}
\renewcommand*{\arraystretch}{1.3}
\Psi(\eta, \mathbf{k}) \equiv
\begin{pmatrix}
\delta(\eta, \mathbf{k}) \\
- \theta(\eta, \mathbf{k}) / ( f(\eta) \, \mathcal{H}(\eta) ) \\
\varsigma(\eta, \mathbf{k}) / ( f(\eta) \, \mathcal{H}(\eta) )^2 \\
\vartheta(\eta, \mathbf{k}) / ( f(\eta) \, \mathcal{H}(\eta) )^2 \\
\omega_i(\eta, \mathbf{k}) / ( f(\eta) \, \mathcal{H}(\eta) ) \\
\vartheta_i(\eta, \mathbf{k}) / ( f(\eta) \, \mathcal{H}(\eta) )^2 \\
\vartheta_{ij}(\eta, \mathbf{k}) / ( f(\eta) \, \mathcal{H}(\eta) )^2
\end{pmatrix} \; .
\label{eq:fluctuation_fields_vector}
\end{equation}
The component fields are referred to by $\Psi_a(\eta, \mathbf{k})$ where the index $a$ runs over all fields and implicitly carries possible spatial indices which are only displayed if necessary. We use the Einstein summation convention which implies summation over possible carried vector or tensor indices. Scalar, vector and tensor fields are referred to by the indices $a_s$, $a_v$ and $a_t$ which run over the scalar, vector and tensor field components only. When only the first two components of $\Psi_a(\eta, \mathbf{k})$ are considered, that is the density and velocity-divergence fields, the analysis reduces to that of perfect pressureless fluid dark matter \cite{scoccimarro_2001, crocce_2006_1, crocce_2006_2}.
\par
Similar to before we define the equal-time power spectra
\begin{equation}
(2 \pi)^3 \, \delta_\text{D}^{(3)}( \mathbf{k} + \mathbf{k}^\prime ) \, P_{ab} (\eta, k) \equiv \left\langle \Psi_a (\eta, \mathbf{k}) \; \Psi_b (\eta, \mathbf{k}^\prime) \right\rangle \; ,
\label{eq:definition_PS}
\end{equation}
which depend only on the wave number $k$ due to the statistical homogeneity and isotropy symmetry. The power spectrum \eqref{eq:definition_PS} is naturally block diagonal for scalar, vector and tensor fields and is understood to implicitly carry a possible projector for the vector and tensor fields, i.e.
\begin{equation}
\begin{aligned}
(2 \pi)^3 \, \delta_\text{D}^{(3)}( \mathbf{k} + \mathbf{k}^\prime ) \, \Delta_{ij}(\mathbf{k}) \, P_{a_v b_v} (\eta, k) &= \left\langle \Psi_{a_v}^i(\eta, \mathbf{k}) \, \Psi_{b_v}^j(\eta, \mathbf{k}^\prime) \right\rangle \; , \\
(2 \pi)^3 \, \delta_\text{D}^{(3)}( \mathbf{k} + \mathbf{k}^\prime ) \, \Delta_{ijkl}(\mathbf{k}) \, P_{a_t b_t} (\eta, k) &= \left\langle \Psi_{a_t}^{ij}(\eta, \mathbf{k}) \, \Psi_{b_t}^{kl}(\eta, \mathbf{k}^\prime) \right\rangle \; .
\end{aligned}
\label{eq:vector_tensor_power_spectra}
\end{equation}
It is convenient to also define the rescaled background
\begin{equation}
\hat{\sigma}(\eta) \equiv \frac{\bar{\sigma}(\eta)}{f^2(\eta) \, \mathcal{H}^2(\eta)} \; ,
\label{eq:rescaled_background}
\end{equation}
and introduce the rescaled source
\begin{equation}
\hat{Q}(\eta) \equiv \frac{1}{3} \int_\mathbf{q} \Big[ - P_{32}(\eta, q) + 2 \, P_{42}(\eta, q) - 4 \, P_{65}(\eta, q) \Big] \; ,
\label{eq:definition_Q_hat}
\end{equation}
such that the evolution equation \eqref{eq:background_velocity_dispersion_eq_v2} can be written as
\begin{equation}
\partial_\eta \hat{\sigma}(\eta) + \left( 3 \, \frac{\Omega_m(\eta)}{f^2(\eta)} - 2 \right) \hat{\sigma}(\eta) - \hat{Q}(\eta) = 0 \; .
\label{eq:background_velocity_dispersion_eq_v3}
\end{equation}
The equations of motion \eqref{eq:continuity_eq_v2} -- \eqref{eq:tensor_velocity_dispersion_eq} can be cast into the form
\begin{equation}
\partial_\eta \Psi_a(\eta, \mathbf{k}) + \Omega_{ab}(\eta, k) \, \Psi_b(\eta, \mathbf{k}) + I_a(\eta, \mathbf{k}) + J_a(\eta, \mathbf{k}) = 0 \; .
\label{eq:eom_v1}
\end{equation}
The coefficients of the terms linear in the fluctuation fields are given by the matrix $\Omega_{ab}(\eta, k)$, which is block diagonal for scalar, vector and tensor fields. The scalar submatrix is given by
\begin{equation}
\renewcommand*{\arraystretch}{1.3}
\Omega_{a_s b_s}(\eta, k) =
\begin{pmatrix}
0 & -1 & 0 & 0 \\
- \frac{3}{2} \frac{\Omega_m(\eta)}{f^2(\eta)} + k^2 \hat{\sigma}(\eta) & \frac{3}{2} \frac{\Omega_m(\eta)}{f^2(\eta)} - 1 & k^2 & k^2 \\
0 & - \frac{2}{3} \, \hat{\sigma}(\eta) & 3 \, \frac{\Omega_m(\eta)}{f^2(\eta)} - 2 & 0 \\
0 & - \frac{4}{3} \, \hat{\sigma}(\eta) & 0 & 3 \, \frac{\Omega_m(\eta)}{f^2(\eta)} - 2
\end{pmatrix} \; ,
\label{eq:omega_scalar}
\end{equation}
while the vector and tensor submatrices are
\begin{equation}
\renewcommand*{\arraystretch}{1.3}
\Omega_{a_v b_v}(\eta, k) =
\begin{pmatrix}
\frac{3}{2} \frac{\Omega_m(\eta)}{f^2(\eta)} - 1 & - k^2 \\
\hat{\sigma}(\eta) & 3 \, \frac{\Omega_m(\eta)}{f^2(\eta)} - 2
\end{pmatrix}
\; , \qquad \Omega_{a_t b_t}(\eta, k) = 3 \, \frac{\Omega_m(\eta)}{f^2(\eta)} - 2 \; .
\label{eq:omega_vector_tensor}
\end{equation}
Similar to the power spectra \eqref{eq:vector_tensor_power_spectra} the submatrices $\Omega_{a_v b_v}(\eta, k)$ and $\Omega_{a_t b_t}(\eta, k)$ are understood to implicitly carry a vector or tensor projector, respectively. The non-linear terms quadratic in the fluctuation fields are given by
\begin{equation}
I_a(\eta, \mathbf{k}) = \int_{\mathbf{k}_1, \mathbf{k}_2} \delta_\text{D}^{(3)}\left( \mathbf{k} - \mathbf{k}_1 - \mathbf{k}_2 \right) \, Y_{abc}(\eta, \mathbf{k}, \mathbf{k}_1, \mathbf{k}_2) \, \Psi_b (\eta, \mathbf{k}_1) \, \Psi_c (\eta, \mathbf{k}_2) \; ,
\end{equation}
whereas the terms cubic in the fluctuation fields are
\begin{equation}
\begin{aligned}
J_a(\eta, \mathbf{k}) = \int_{\mathbf{k}_1, \mathbf{k}_2, \mathbf{k}_3} \delta_\text{D}^{(3)}\left( \mathbf{k} - \mathbf{k}_1 - \mathbf{k}_2 - \mathbf{k}_3 \right) \, X_{abcd}&(\eta, \mathbf{k}, \mathbf{k}_1, \mathbf{k}_2, \mathbf{k}_3) \\
&\times \Psi_b (\eta, \mathbf{k}_1) \, \Psi_c (\eta, \mathbf{k}_2) \, \Psi_d (\eta, \mathbf{k}_3) \; .
\end{aligned}
\end{equation}
Here $Y_{abc}(\eta, \mathbf{k}, \mathbf{k}_1, \mathbf{k}_2)$ and $X_{abcd}(\eta, \mathbf{k}, \mathbf{k}_1, \mathbf{k}_2, \mathbf{k}_3)$ are symmetrised versions of the vertices appearing in equations \eqref{eq:quadratic_terms_continuity_eq} -- \eqref{eq:cubic_terms_vorticity_eq}.

\section{Evolution of the background and linear fluctuations}
\label{sec:evolution_of_the_background_and_linear_fluctuations}
\subsection{Linear response}
\label{sec:linear_equations_of_motion}
At early times and large scales we assume the fields $\Psi_a(\eta, \mathbf{k})$ to be small compared to the homogeneous and isotropic background. Therefore, it seems natural to study how the fields evolve in the linear regime where we neglect the coupling of different Fourier modes. In the limit where we turn off interactions, the equations of motion \eqref{eq:eom_v1} reduce to
\begin{equation}
\partial_\eta \Psi_a(\eta, \mathbf{k}) + \Omega_{ab}(\eta, k) \, \Psi_b(\eta, \mathbf{k}) = 0 \; .
\label{eq:linear_eom}
\end{equation}
While the equations of motion \eqref{eq:linear_eom} describe linear fields, in the sense that no interaction between Fourier modes is taken into account, they nevertheless form a set of non-linear differential equations. Because the velocity dispersion background $\hat{\sigma}(\eta)$ enters $\Omega_{ab}(\eta, k)$ the evolution equation \eqref{eq:background_velocity_dispersion_eq_v3} is needed to close the system of equations, which in turn depends on the correlation of the fluctuation fields through the source $\hat{Q}(\eta)$. In this situation one cannot expect to find solutions in a closed form in terms of standard mathematical functions. 
\par
To fully determine the system we need to specify initial conditions at some initial time evolution parameter $\etain \equiv \eta(\tau_\text{in})$.\footnote{By definition \eqref{eq:time_evolution_parameter} the initial time evolution parameter is $\etain \equiv 0$.} In the following we use the approximation $\Omega_m(\tau) / f^2(\tau) \approx 1$, which is sufficiently accurate for the times we are interested in \cite{scoccimarro_1998}. For an initial velocity dispersion background $\hat{\sigma}_\text{in}$, the evolution equation \eqref{eq:background_velocity_dispersion_eq_v3} can be solved with
\begin{equation}
\hat{\sigma}(\eta) = \hat{\sigma}_\text{in} \, e^{- (\eta - \etain)} + e^{- (\eta - \etain)} \int_{\etain}^\eta d\xi \, e^{\xi - \etain} \, \hat{Q}(\xi) \; .
\label{eq:velocity_dispersion_background_solution}
\end{equation}
The fields $\Psi_a(\eta, \mathbf{k})$ can formally be written as a linear response to the initial conditions $\Phi_a(\mathbf{k})$,
\begin{equation}
\Psi_a(\eta, \mathbf{k}) = \gR{ab}(\eta, \etain, k) \, \Phi_b(\mathbf{k}) \; .
\label{eq:linear_impulse_response}
\end{equation}
The response function $\gR{ab}(\eta, \etain, k)$ is subject to the defining equation
\begin{equation}
\Big[ \partial_\eta \delta_{ab} + \Omega_{ab}(\eta, k) \Big] \, g_{bc}^R(\eta, \etain, k) = \delta_{ac} \, \delta_\text{D}^{(1)}(\eta - \etain) \; ,
\label{eq:retarded_linear_propagator_equation}
\end{equation}
together with the causal boundary conditions
\begin{equation}
\begin{aligned}
&\gR{ab}(\eta, \etain, k) = 0 &&\text{for} \; \eta < \etain \; , \\
&\gR{ab}(\eta, \etain, k) \to \delta_{ab} &&\text{for} \; \eta \to \etain^+ \; ,
\end{aligned}
\end{equation}
and is henceforth called the retarded linear propagator. With the initial power spectrum
\begin{equation}
(2 \pi)^3 \, \delta_\text{D}^{(3)}( \mathbf{k} + \mathbf{k}^\prime ) \, \Pin{ab}(k) = \left\langle \Phi_a(\mathbf{k}) \, \Phi_b(\mathbf{k}^\prime) \right\rangle \; ,
\end{equation}
the source \eqref{eq:definition_Q_hat} can be written as
\begin{equation}
\begin{aligned}
\hat{Q}(\eta) = \frac{4 \pi}{3} \int_0^\infty \frac{dq \, q^2}{(2 \pi)^3} \, \bigg\{ \Big[ - \gR{3a}(\eta, \etain, q) + 2 \, &\gR{4a}(\eta, \etain, q) \Big] \, \gR{2b}(\eta, \etain, q) \\
&- \gR{6a}(\eta, \etain, q) \, \gR{5b}(\eta, \etain, q) \bigg\} \, \Pin{ab}(q) \; .
\label{eq:R_function_retarded_linear_propagator}
\end{aligned}
\end{equation}
We emphasise that equation \eqref{eq:retarded_linear_propagator_equation} is non-linear and the retarded linear propagator is not a Green's function in the standard sense of linear differential equations, because $g_{ab}^R(\eta, \etain, k)$ depends on the initial velocity dispersion background $\hat{\sigma}_\text{in}$ and is a functional of the initial power spectrum $\Pin{ab}(k)$. As a consequence, the linear fields do not grow independently for different Fourier modes in contrast to the single-stream approximation, because $\hat{Q}(\eta)$ is sensitive to the evolution of the fluctuation fields for all wave numbers $k$. Nevertheless, equation \eqref{eq:linear_impulse_response} can be seen as a \emph{linear} response to a \emph{small} perturbation in the initial random field.

\subsection{Initial conditions}
In the following we assume that the initial vector and tensor fields can be neglected and that the initial scalar fields can be written as proportional random fields,
\begin{equation}
\Phi_{a_s}(\mathbf{k}) = w_{a_s}(k) \, \delta^\text{in}(\mathbf{k}) \; ,
\label{eq:scalar_fluctuation_field_initial_conditions}
\end{equation}
where $\delta^\text{in}(\mathbf{k})$ is the initial dark matter mass density fluctuation field. The vector $w_{a_s}(k)$ quantifies the proportionality of the initial scalar fields to $\delta^\text{in}(\mathbf{k})$ and we chose it to be the eigenvector of the growing solution given in section \ref{sec:scalar_growth_factors_and_free-streaming_scale}. Further, we assume $\delta^\text{in}(\mathbf{k})$ to be a Gaussian random field which is completely characterised by the initial scalar power spectrum
\begin{equation}
(2 \pi)^3 \, P_s^\text{in}(k) \, \delta_\text{D}^{(3)}(\mathbf{k} + \mathbf{k}^\prime) \equiv \left\langle \delta_\text{in}(\mathbf{k}) \, \delta_\text{in}(\mathbf{k}^\prime) \right\rangle .
\end{equation}
It is particularly convenient to discuss solutions in terms of the linear growth functions
\begin{equation}
g_a^R(\eta, k) \equiv \gR{ab}(\eta, \etain, k) \, w_b(k) \; ,
\label{eq:linear_growth_functions}
\end{equation}
such that the equal-time linear evolved scalar power spectra can be written as
\begin{equation}
P_{a_s b_s}(\eta, k) = g_{a_s}^R(\eta, k) \, g_{b_s}^R(\eta, k) \, P_s^\text{in}(k) \; .
\end{equation}
For a vanishing initial vector power spectrum the source for the velocity dispersion background takes the form
\begin{equation}
\hat{Q}(\eta) = \frac{4 \pi}{3} \int_0^\infty \frac{dq \, q^2}{(2 \pi)^3} \, \Big[ - \gR{3}(\eta, q) + 2 \, \gR{4}(\eta, q) \Big] \, \gR{2}(\eta, q) \, \Pin{s}(q) \; ,
\label{eq:Q_hat_scalar_initial_conditions}
\end{equation}
which depends on the scalar linear growth functions only.
\par
Since $\Omega_{ab}(\eta, k)$ is block diagonal for scalar, vector and tensor fields, and because $\hat{Q}(\eta)$ depends on the scalar linear growth functions only, we can solve the scalar part of equation \eqref{eq:retarded_linear_propagator_equation} together with the velocity dispersion background \eqref{eq:velocity_dispersion_background_solution} independent from the vector and tensor parts. Subsequently the vector and tensor part can be solved.

\subsection{Scalar growth factors and free-streaming scale}
\label{sec:scalar_growth_factors_and_free-streaming_scale}
Because the scalar retarded linear propagator is a functional of the initial power spectrum, we need to specify $w_{a_s}(k)$ in order to solve the scalar part of equation \eqref{eq:retarded_linear_propagator_equation}. To make a sensible choice we consider the eigenvalues of the submatrix $- \Omega_{a_s b_s}(\eta, k)$ which determine the growth of the solutions at time $\eta$ and wave number $k$ and are henceforth called growth factors. Three growth factors are given by the roots of the polynomial
\begin{equation}
\Big[ s_s(\eta, k) - 1 \Big] \bigg[ s_s(\eta, k) + \frac{3}{2} \bigg] \Big[ s_s(\eta, k) + 1 \Big] + k^2 \, \hat{\sigma}(\eta) \Big[ 3 \, s_s(\eta, k) + 1 \Big] \; ,
\label{eq:scalar_growth_factor_polynomial}
\end{equation}
whereas the fourth is $s_s(\eta, k) = - 1$. In the limit $k \to 0$, one can easily find the roots of the polynomial \eqref{eq:scalar_growth_factor_polynomial} and obtains
\begin{equation}
\begin{aligned}
s_s^{(1)}(\eta, k) &= 1 + \mathcal{O}(k^2) \; , \qquad &&s_s^{(2)}(\eta, k) = - \frac{3}{2} + \mathcal{O}(k^2) \; , \\
s_s^{(3)}(\eta, k) &= - 1 + \mathcal{O}(k^2) \; , \qquad &&s_s^{(4)}(\eta, k) = - 1 \; .
\end{aligned}
\end{equation}
To fully characterise the solutions the corresponding eigenvectors are needed. These are given by
\begin{equation}
v_{a_s}^{(n)}(\eta, k) =
\begin{pmatrix}
1 \\
s_s^{(n)}(\eta, k) \\[0.7ex]
\dfrac{2 \, \hat{\sigma}(\eta) \, s_s^{(n)}(\eta, k)}{3 \, (1 + s_s^{(n)}(\eta, k))} \\[2.7ex]
\dfrac{4 \, \hat{\sigma}(\eta) \, s_s^{(n)}(\eta, k)}{3 \, (1 + s_s^{(n)}(\eta, k))}
\end{pmatrix}
\; , \qquad
\renewcommand*{\arraystretch}{1.0}
v_{a_s}^{(4)}(\eta, k) =
\begin{pmatrix}
0 \\ 0 \\ -1 \\ 1
\end{pmatrix}
\; ,
\label{eq:scalar_eigenvectors}
\end{equation}
for $n \in \{ 1, 2, 3 \}$.
\par
In the limit $k \to 0$ the density and velocity-divergence fields decouple from the velocity dispersion fields. In turn the former evolve according to the dynamics of the single-stream approximation and obey a growing and decaying solution proportional to $e^{\eta - \etain}$ and $e^{- \frac{3}{2} (\eta - \etain)}$ \cite{scoccimarro_2001} while the latter decay proportional to $e^{- (\eta - \etain)}$. The growth factors $s_s^{(1)}(\eta, 0)$ and $s_s^{(2)}(\eta, 0)$ can be identified with the single-stream approximation solutions, whereas $s_s^{(3)}(\eta, 0)$ and $s_s^{(4)}(\eta, 0)$ correspond to the two new decaying solutions.
\par
\begin{figure}[ht]
\centering
\includegraphics[scale=0.5]{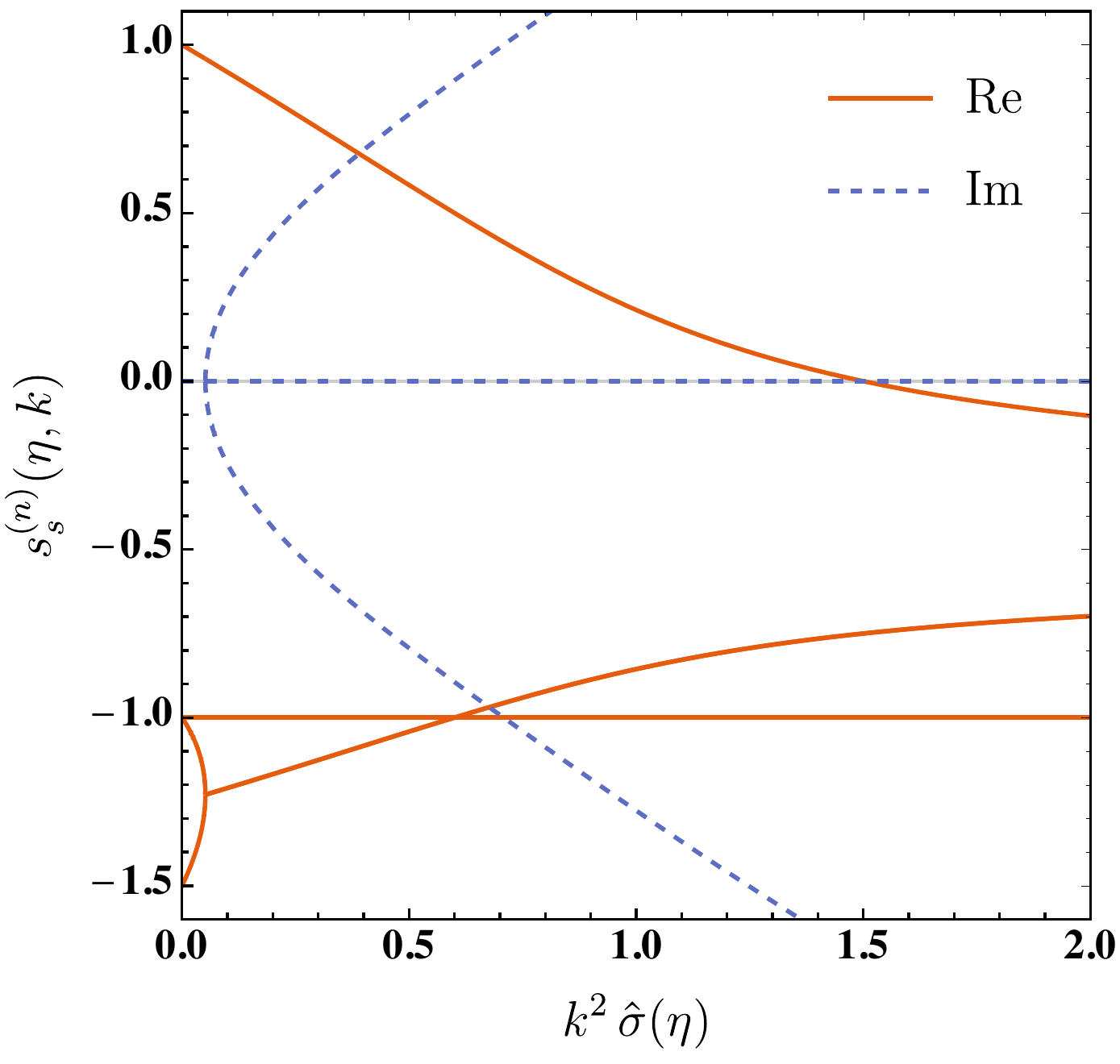}
\caption{Scalar field growth factors $s_s^{(n)}(\eta, k)$ as a function of the combination $k^2 \, \hat{\sigma}(\eta)$. The growth factors seem to admit only a single growing solution in the range $0 \le k^2 \, \hat{\sigma}(\eta) < 3 / 2$. See text for further discussion.}
\label{fig:scalar_field_growth_factors}
\end{figure}
For finite values of $k^2 \, \hat{\sigma}(\eta)$ the picture is more complicated. In figure \ref{fig:scalar_field_growth_factors} we show the real and imaginary parts of the scalar field growth factors as a function of the combination $k^2 \, \hat{\sigma}(\eta)$. The natural extension of the single-stream approximation growth factor $s_s^{(1)}(\eta, k)$ is the only positive growth factor in the range $0 \le k^2 \, \hat{\sigma}(\eta) < 3 / 2$ and turns negative for larger values of $k^2 \, \hat{\sigma}(\eta)$. While $s_s^{(1)}(\eta, k)$ and $s_s^{(4)}(\eta, k)$ are real for all $k^2 \, \hat{\sigma}(\eta)$ the growth factors $s_s^{(2)}(\eta, k)$ and $s_s^{(3)}(\eta, k)$ develop an imaginary part at some $k^2 \, \hat{\sigma}(\eta) > 0$ from where on their real parts stay negative and coincide.
\par
We emphasis that due to the time dependence of $\Omega_{ab}(\eta, k)$ the retarded linear propagator is not time translation invariant and thus the eigenvectors generally differ from those of $\Omega_{ab}(\eta, k)$. Therefore exciting a solution proportional to an eigenvector \eqref{eq:scalar_eigenvectors} at initial time $\etain$ naturally evolves to a superposition of different eigenvectors at time $\eta$. Further, the full time dependence of a solution is determined by the growth factors as well as their corresponding eigenvectors.
\par
In the following we concentrate on the growing solution characterised by the growth factor $s_s^{(1)}(\eta, k)$. We assume that initially only the growing solution is present and the corresponding eigenvector determines the initial scalar field configuration, i.e. 
\begin{equation}
w_{a_s}(k) = v_{a_s}^{(1)}(\etain, k) \; .
\label{eq:initial_scalar_field_configuration}
\end{equation}
Before solving the system of equations \eqref{eq:velocity_dispersion_background_solution} and \eqref{eq:retarded_linear_propagator_equation} by numerical means, it is sensible to infer some properties of the solution. The growth factor $s_s^{(1)}(\eta, k)$ is a monotonically decreasing function of $k$ and has a zero-crossing at the comoving free-streaming wave number
\begin{equation}
k_\text{fs}(\eta) \equiv \sqrt{\frac{3}{2 \, \hat{\sigma}(\eta)}} \; ,
\label{eq:free-streaming_wave number}
\end{equation}
and is negative for all $k > k_\text{fs}(\eta)$. The wave number $k_\text{fs}(\eta)$ determines the scale below which the fluctuation fields do not grow due to the free-streaming of dark matter particles. The corresponding solution is therefore damped at smaller scales and decays for wave numbers larger than $k_\text{fs}(\eta)$. In order to qualitatively understand how $k_\text{fs}(\eta)$ evolves in time, we consider the limit of early and late times. At early times the fluctuation fields are small in amplitude and one can neglect the enhancement of the velocity dispersion background \eqref{eq:velocity_dispersion_background_solution} due to $\hat{Q}(\eta)$ as long as the variance of the velocity-divergence and scalar velocity dispersion fields is smaller than the velocity dispersion background,
\begin{equation}
\hat{Q}(\eta) \ll \hat{\sigma}(\eta) \; .
\label{eq:background_velocity_dispersion_early_time_limit}
\end{equation}
In this limit the velocity dispersion background decays in time and in turn the free-streaming wave number is raised which allows structures to form at smaller scales. At later times, however, the fluctuation fields grow in amplitude and the validity of the limit \eqref{eq:background_velocity_dispersion_early_time_limit} breaks down. Eventually, at times where
\begin{equation}
\hat{Q}(\eta) \gtrsim \hat{\sigma}(\eta) \; ,
\label{eq:background_velocity_dispersion_late_time_limit}
\end{equation}
the velocity dispersion background starts to grow and in turn the free-streaming wave number is lowered which stalls the formation of structure at small scales.

\subsection{Vector and tensor growth factors}
\label{sec:vector_and_tensor_growth_factors}
Similar to the case of scalar fields it is sensible to study properties of the vector retarded linear propagator in terms of the eigenvalues of the submatrix $- \Omega_{a_v b_v}(\eta, k)$. The growth factors for the vector fields are given by
\begin{equation}
\begin{aligned}
s_v^{(1)}(\eta, k) &= - \frac{3}{4} + \sqrt{ \frac{1}{16} - k^2 \, \hat{\sigma}(\eta)} \; , \\
s_v^{(2)}(\eta, k) &= - \frac{3}{4} - \sqrt{ \frac{1}{16} - k^2 \, \hat{\sigma}(\eta)} \; ,
\end{aligned}
\label{eq:vector_growth_factors}
\end{equation}
and the corresponding eigenvectors are
\begin{equation}
v_{a_v}^{(n)}(\eta, k) =
\begin{pmatrix}
- 1 - s_v^{(n)}(\eta, k) \\  \hat{\sigma}(\eta)
\end{pmatrix}
\; ,
\end{equation}
for $n \in \{ 1, 2 \}$. In the limit $k \to 0$ one obtains
\begin{equation}
s_v^{(1)}(\eta, k) = - \frac{1}{2} + \mathcal{O}(k^2) \; , \qquad s_v^{(2)}(\eta, k) = - 1 + \mathcal{O}(k^2) \; .
\end{equation}
Since the limit $k \to 0$ restores the equations of motion of perfect fluid dark matter by decoupling the vorticity field from the vector velocity dispersion field, one can identify $s_v^{(1)}(\eta, 0)$ with the decaying solution of the single-stream approximation vorticity field proportional to $e^{- \frac{1}{2} (\eta - \etain)}$. The growth factor $s_v^{(2)}(\eta, 0)$ corresponds to a new decaying solution proportional to $e^{- (\eta - \etain)}$, similar to the scalar case.
\par
\begin{figure}[ht]
\centering
\includegraphics[scale=0.5]{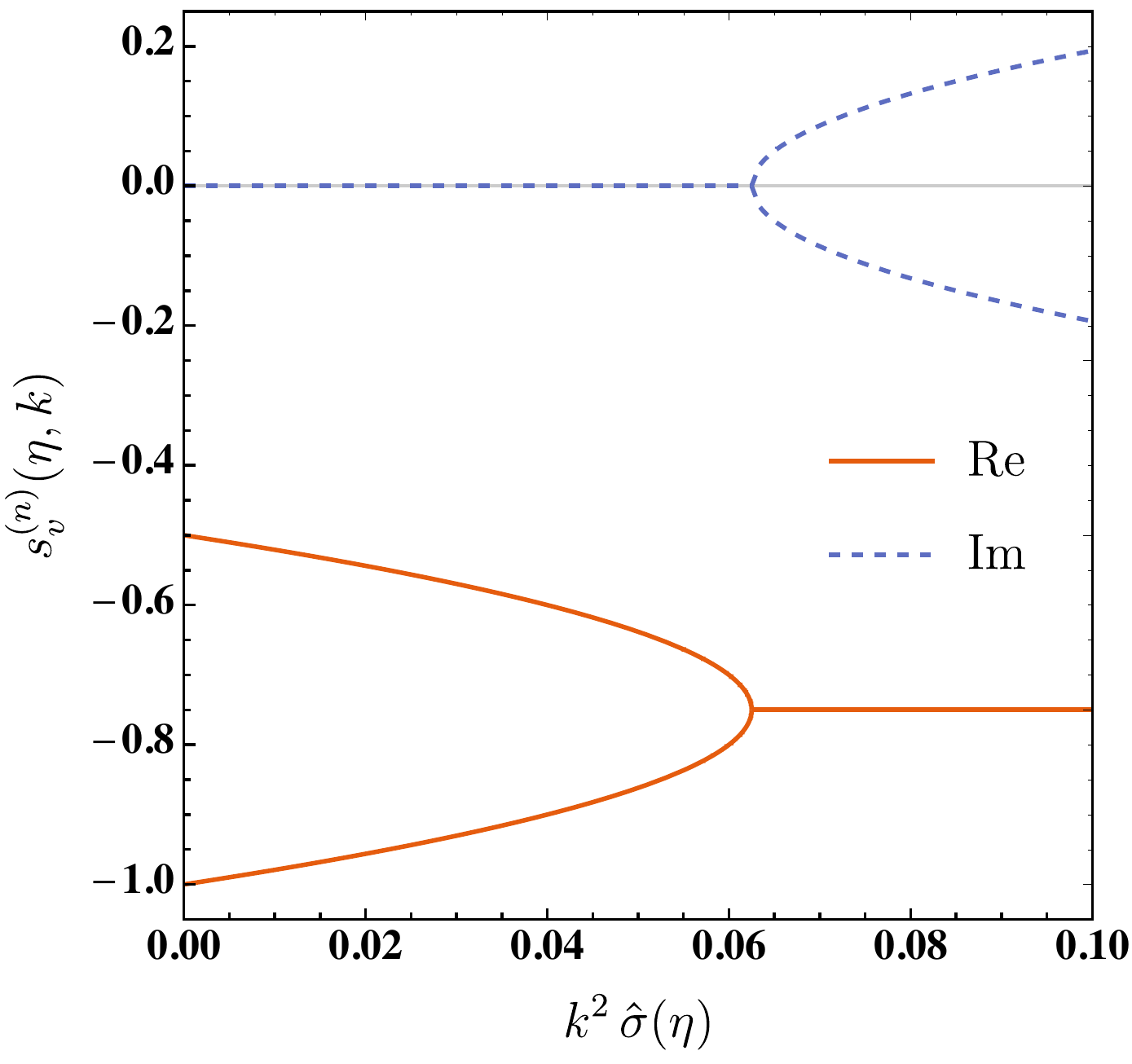}
\caption{Vector field growth factors $s_v^{(n)}(\eta, k)$ as a function of the combination $k^2 \, \hat{\sigma}(\eta)$. The growth factors seem to admit only decaying solutions. See text for further discussion.}
\label{fig:vector_field_growth_factors}
\end{figure}
For finite values of $k^2 \, \hat{\sigma}(\eta)$, the real and imaginary parts of the vector field growth factors are shown in figure \ref{fig:vector_field_growth_factors}. Both growth factors $s_v^{(1)}(\eta, k)$ and $s_v^{(2)}(\eta, k)$ are real and negative for sufficient small $k^2 \, \hat{\sigma}(\eta)$. For larger $k^2 \, \hat{\sigma}(\eta)$ they develop an imaginary part from where on their real parts coincide and take a constant negative value. The growth factor $s_v^{(1)}(\eta, k)$ is monotonically decreasing while $s_v^{(2)}(\eta, k)$ is monotonically increasing for larger $k^2 \, \hat{\sigma}(\eta)$ until they develop an imaginary part.
\par
Because the vector part of equation \eqref{eq:retarded_linear_propagator_equation} is a first-order linear ordinary differential equation, any general solution can be constructed by superposition of the solutions characterised by the growth factors $s_v^{(n)}(\eta, k)$ for $n \in \{ 1, 2 \}$.\footnote{This is only true as long as we neglect initial vector fields which would enter the source of the velocity dispersion background by definition \eqref{eq:definition_Q_hat} and thus couple the scalar and vector part of equation \eqref{eq:retarded_linear_propagator_equation}.} The corresponding initial field configurations are determined by the eigenvectors, i.e.
\begin{equation}
w_{a_v}^{(n)}(k) = v_{a_v}^{(n)}(\etain, k) \; .
\end{equation}
Finally, the tensor part of equation \eqref{eq:retarded_linear_propagator_equation} together with the submatrix \eqref{eq:omega_vector_tensor} is independent of $k$ and the tensor retarded linear propagator can therefore be given in the closed form
\begin{equation}
g_{a_t b_t}^R(\eta, \eta^\prime, k) = e^{- (\eta - \eta^\prime)} \; .
\end{equation}
The corresponding tensor growth factor is $s_t(\eta, k) = - 1$ for which we simply chose $w_{a_t}(k) = \hat{\sigma}_\text{in}$ as the initial tensor field configuration.

\subsection{Approximation for small velocity dispersion and soft power spectrum}
\label{sec:approximation_for_small_velocity_dispersion_and_soft_power_spectrum}
In the following we discuss an approximate analytical solution that becomes available in the limit where the velocity dispersion is small and where the initial scalar power spectrum $P_s^\text{in}(k)$ does not extend too far into the ultraviolet, in a sense that becomes clear below.
\par
Assuming that the velocity dispersion background and fluctuation fields are small, we consider the equations \eqref{eq:retarded_linear_propagator_equation} at wave numbers $k^2 \, \hat{\sigma}(\eta) \ll 1$, $k^2 \, \varsigma(\eta) \ll 1$, $k^2 \, \vartheta(\eta) \ll 1$ and $k^2 \, \vartheta_i(\eta) \ll 1$. Since the evolution of the velocity dispersion background depends on $\hat{Q}(\eta)$, which involves an integral over all wave numbers, this assumption does on first sight not seem sensible. However, as discussed in section \ref{sec:numerical_results}, the initial scalar power spectrum $P_s^\text{in}(k)$ is heavily suppressed for wave numbers larger than the comoving free-streaming wave number at matter-radiation equality $\kfseq$. The above proposed approximation can therefore be used as long as $\kfseq^2 \, \hat{\sigma}(\eta) \ll 1$, $\kfseq^2 \, \varsigma(\eta) \ll 1$, $\kfseq^2 \, \vartheta(\eta) \ll 1$ and $\kfseq^2 \, \vartheta_i(\eta) \ll 1$. 
\par
By dropping the $k$ dependent terms in equations \eqref{eq:retarded_linear_propagator_equation} and denoting objects in this approximation by a tilde, the equations of motion for the linear growth functions can be written as
\begin{equation}
\Big[ \partial_\eta \delta_{ab} + \tilde{\Omega}_{ab}(\eta) \Big] \, \tilde{g}_b^R(\eta) = w_a(0) \, \delta_\text{D}^{(1)}(\eta - \etain) \; ,
\label{eq:approximation_eom}
\end{equation}
where $\tilde{\Omega}_{ab}(\eta)$ is given by the submatrices
\begin{equation}
\renewcommand*{\arraystretch}{1.3}
\tilde{\Omega}_{a_s b_s}(\eta) \equiv
\begin{pmatrix}
0 & -1 & 0 & 0 \\
- \frac{3}{2} & \frac{1}{2} & 0 & 0 \\
0 & - \frac{2}{3} \, \hat{\sigma}(\eta) & 1 & 0 \\
0 & - \frac{4}{3} \, \hat{\sigma}(\eta) & 0 & 1
\end{pmatrix}
\; , \quad
\tilde{\Omega}_{a_v b_v}(\eta) \equiv
\begin{pmatrix}
\frac{1}{2} & 0 \\
\hat{\sigma}(\eta) & 1
\end{pmatrix}
\; , \quad
\tilde{\Omega}_{a_t b_t}(\eta) \equiv 1 \; .
\end{equation}
The source function \eqref{eq:Q_hat_scalar_initial_conditions} takes the form
\begin{equation}
\tilde{Q}(\eta) = \frac{\sigma_d^2}{3} \, \Big[ - \tilde{g}_3^R(\eta) + 2 \, \tilde{g}_4^R(\eta) \Big] \, \tilde{g}_2^R(\eta) \; ,
\label{eq:approximation_R}
\end{equation}
where
\begin{equation}
\sigma_d^2 \equiv 4 \pi \int_0^\infty \frac{dq \, q^2}{(2 \pi)^3} \, P_s^\text{in}(q) \; ,
\label{eq:definition_sigma_d}
\end{equation}
is the dimensionless variance of the initial density fluctuation field and is the only quantity appearing in the approximation which is sensitive to the initial scalar power spectrum.\footnote{Notice the difference to the single-stream approximation which depends even at one-loop level only on the dimensionful parameter $\sigma_v^2 \equiv 4 \pi/3 \int_0^\infty dq \, P_s^\text{in}(q)/(2 \pi)^3$ \cite{blas_2013_2}. For a perfectly cold dark matter power spectrum as in equation \eqref{eq:CDM_power_spectrum}, the integral \eqref{eq:definition_sigma_d} diverges approximately logarithmically.}
\par
The initial conditions are taken to be the $k \to 0$ limit of the initial field configuration $w_a(k)$ specified in section \ref{sec:scalar_growth_factors_and_free-streaming_scale} and \ref{sec:vector_and_tensor_growth_factors}. For scalar fields excited in the growing solution with $s_s^{(1)}(\eta, 0) = 1$ these are given by
\begin{equation}
w_{a_s}(0) =
\begin{pmatrix}
1 \\ 1 \\ \dfrac{\hat{\sigma}_\text{in}}{3} \\[1.8ex] \dfrac{2 \, \hat{\sigma}_\text{in}}{3}
\end{pmatrix}
\; .
\label{eq:approximation_scalar_initial_field_configuration}
\end{equation}
The vector fields corresponding to $s_v^{(1)}(\eta, 0) = - 1/2$ and $s_v^{(2)}(\eta, 0) = - 1$ are excited by
\begin{equation}
w_{a_v}^{(1)}(0) =
\begin{pmatrix}
- \frac{1}{2} \\ \hat{\sigma}_\text{in}
\end{pmatrix}
\; , \qquad
w_{a_v}^{(2)}(0) =
\begin{pmatrix}
0 \\ \hat{\sigma}_\text{in}
\end{pmatrix}
\; ,
\end{equation}
respectively and for the tensor field corresponding to $s_t(\eta, 0) = - 1$ we use $w_{a_t}(0) = \hat{\sigma}_\text{in}$. 
\par
To close the system of equations \eqref{eq:approximation_eom} the velocity dispersion background
\begin{equation}
\tilde{\sigma}(\eta) = \hat{\sigma}_\text{in} \, e^{- (\eta - \etain)} + e^{- (\eta - \etain)} \int_{\etain}^\eta d\xi \, e^{\xi - \etain} \, \tilde{Q}(\xi) \; .
\end{equation}
is needed.
\par
The linear growth functions of the density and velocity-divergence fields obey the growing solution
\begin{equation}
\tilde{g}_1^R(\eta) = e^{\eta - \etain} \; , \qquad \tilde{g}_2^R(\eta) = e^{\eta - \etain} \; ,
\label{eq:analytical_scalar_linear_growth_function_12}
\end{equation}
while the scalar velocity dispersion linear growth functions are
\begin{equation}
\tilde{g}_3^R(\eta) = \frac{\hat{\sigma}_\text{in}}{3} \, e^{- (\eta - \etain)} \, H_1(\eta) \; , \qquad \tilde{g}_4^R(\eta) = \frac{2 \, \hat{\sigma}_\text{in}}{3} \, e^{- (\eta - \etain)} \, H_1(\eta) \; .
\label{eq:analytical_scalar_linear_growth_function_34}
\end{equation}
Here we defined the function
\begin{equation}
H_1(\eta) \equiv \frac{2}{C} \, \sinh( C \, (e^{\eta - \etain} - 1)) + \cosh( C \, (e^{\eta - \etain} - 1)) \; ,
\end{equation}
where $C \equiv \sqrt{ 2 \, \sigma_d^2 / 3}$ is a real positive constant. The vorticity field linear growth function is given by
\begin{equation}
\tilde{g}_5^R(\eta) = w_5^{(n)}(0) \, e^{- \frac{1}{2} (\eta - \etain)} \; ,
\label{eq:analytical_vector_growth_function_5}
\end{equation}
and the vector velocity dispersion linear growth function is
\begin{equation}
\tilde{g}_6^{R}(\eta) = w_6^{(n)}(0) \, e^{- (\eta - \etain)} + 2 \, \hat{\sigma}_\text{in} \, w_5^{(n)}(0) \, e^{- (\eta - \etain)} \, \Big[ H_2(\eta) \, e^{- \frac{1}{2} (\eta - \etain)} + H_3(\eta) - 1 \Big] \; ,
\label{eq:analytical_vector_growth_function_6}
\end{equation}
for $n \in \{ 1, 2 \}$. Here we introduced the functions
\begin{equation}
H_2(\eta) \equiv \frac{C}{2} \, \sinh( C \, (e^{\eta - \etain} - 1)) + \cosh( C \, (e^{\eta - \etain} - 1)) \; ,
\end{equation}
and
\begin{equation}
\begin{aligned}
H_3(\eta) &\equiv \sqrt{C} \left( 1 - \frac{C}{2} \right) e^C \, \bigg[ \frac{\sqrt{\pi}}{2} \, \text{erf}\Big( \sqrt{C} \; e^{\frac{1}{2} (\eta - \etain)} \Big) - \frac{\sqrt{\pi}}{2} \, \text{erf}\Big( \sqrt{C} \Big) \bigg] \\
&\hspace*{0.7cm} - \sqrt{C} \left( 1 + \frac{C}{2} \right) e^{- C} \, \bigg[ \frac{\sqrt{\pi}}{2} \, \text{erfi}\Big( \sqrt{C} \; e^{\frac{1}{2} (\eta - \etain)} \Big) - \frac{\sqrt{\pi}}{2} \, \text{erfi}\Big( \sqrt{C} \Big) \bigg] \; ,
\end{aligned}
\end{equation}
where $\text{erf}(x)$ and $\text{erfi}(x)$ denote the real and imaginary error function, respectively. The tensor velocity dispersion linear growth function has the solution
\begin{equation}
\tilde{g}_7^R(\eta) = w_7(0) \, e^{- (\eta - \etain)} \; ,
\label{eq:analytical_tensor_growth_function}
\end{equation}
which coincides with $\gRtilde{6}(\eta)$ for $n = 2$. Finally, the velocity dispersion background is given by
\begin{equation}
\tilde{\sigma}(\eta) = \hat{\sigma}_\text{in} \, e^{-(\eta - \etain)} \, H_2(\eta) \; .
\label{eq:analytical_background_velocity_dispersion}
\end{equation}
With the analytical solutions \eqref{eq:analytical_scalar_linear_growth_function_12}, \eqref{eq:analytical_scalar_linear_growth_function_34}, \eqref{eq:analytical_vector_growth_function_5}, \eqref{eq:analytical_vector_growth_function_6}, \eqref{eq:analytical_tensor_growth_function} and \eqref{eq:analytical_background_velocity_dispersion} we can infer the early and late time behaviour of the linear growth functions and velocity dispersion background in the approximative scenario.
\par
As for the single-stream approximation, the density and velocity-divergence fields grow at all times exponentially with the time evolution parameter $\eta$ while the vorticity field decays proportional to $e^{- \frac{1}{2} (\eta - \etain)}$. The evolution of the velocity dispersion fields strongly depends on the value of $C$, but since we assume the power spectrum to not extent too far into the ultraviolet, we study the early and late time behaviour for $C \ll 1$. The scalar velocity dispersion fields grow at early times exponentially,
\begin{equation}
\gRtilde{3}(\eta) \propto e^{\eta - \etain} \; , \qquad \gRtilde{4}(\eta) \propto e^{\eta - \etain} \; ,
\label{eq:analytical_scalar_linear_growth_function_3_4_early_times}
\end{equation}
and growth with double exponential power at late times
\begin{equation}
\gRtilde{3}(\eta) \propto \exp \Big\{ C \, e^{\eta - \etain} - (\eta - \etain) \Big\} \; , \qquad \gRtilde{4}(\eta) \propto \exp \Big\{ C \, e^{\eta - \etain} - (\eta - \etain) \Big\} \; .
\label{eq:analytical_scalar_linear_growth_function_3_late_times}
\end{equation}
\par
According to the definitions \eqref{eq:peculiar_velocity_dispersion_tensor_field}, \eqref{eq:scalar_isotropic_velocity_dispersion_fluctuation_field}, \eqref{eq:scalar_anisotropic_velocity_dispersion_fluctuation_field}, \eqref{eq:vector_velocity_dispersion_field}, \eqref{eq:tensor_velocity_dispersion_field} and \eqref{eq:fluctuation_fields_vector} the velocity dispersion linear growth functions describe perturbations relative to the velocity dispersion background. The latter is therefore needed for a complete picture. According to the approximation \eqref{eq:analytical_background_velocity_dispersion} it decays exponentially at early times,
\begin{equation}
\tilde{\sigma}(\eta) \propto e^{-(\eta - \etain)} \; ,
\end{equation}
while at late times one finds in the formal limit $\eta \to \infty$ the dominant behaviour to be a double exponential growth,
\begin{equation}
\tilde{\sigma}(\eta) \propto \exp \Big\{ C \, e^{\eta - \etain} - (\eta - \etain) \Big\} \; .
\label{eq:approximation_velocity_dispersion_background_late_time_limit}
\end{equation}
This qualitatively agrees with the early and late times regimes \eqref{eq:background_velocity_dispersion_early_time_limit} and \eqref{eq:background_velocity_dispersion_late_time_limit} discussed in section \ref{sec:scalar_growth_factors_and_free-streaming_scale}.

\section{Numerical results}
\label{sec:numerical_results}
\begin{table}[H]
\centering
\renewcommand{\arraystretch}{1.2}
\begin{tabular}{ | C{1.9cm} | C{1.9cm} | C{1.9cm} | C{1.9cm} | C{1.9cm} | C{1.9cm} | }
\hline
$h$ & $\Omega_{c,0} \, h^2$ & $\Omega_{b,0} \, h^2$ & $z_\text{re}$ & $A_s$ & $n_s$ \\ \hline
$0.6774$ &  $0.1188$ & $0.0223$ &  $8.8$ & $2.142 \cdot 10^{-9}$ & $0.9667$ \\
\hline
\end{tabular}
\caption{Cosmological parameters found by the Planck Collaboration \cite{planck_2016} where $h$ is the reduced Hubble constant, $\Omega_{c,0}$ and $\Omega_{b,0}$ are the current cold dark matter and baryonic matter density parameters respectively, $z_\text{re}$ is the redshift at reionisation, $A_s$ is the primordial scalar amplitude at the pivot scale $k_* = 0.05 \, h /$Mpc and $n_s$ is the scalar spectral index.}
\label{table:cosmological_parameters}
\end{table}
\subsection{Initial scalar power spectrum and velocity dispersion background}
\label{sec:initial_scalar_power_spectrum_and_background_velocity_dispersion}
To solve the system of equations \eqref{eq:velocity_dispersion_background_solution} and \eqref{eq:retarded_linear_propagator_equation} numerically we need to specify the velocity dispersion background $\sigmahatin$ as well as the scalar power spectrum $P_s^\text{in}(k)$ at an initial time $\tau_\text{in}$. We set $\tau_\text{in}$ deep within the matter dominated era of the Universe at a redshift of $z(\tau_\text{in}) \equiv 1/a(\tau_\text{in}) - 1 = 100$ and assume that the non-linear evolution of perturbations is negligible at earlier times, such that we can use a linearly evolved power spectrum as initial condition. Further we assume that the matter fluctuation fields are initially small enough in amplitude to be in the regime \eqref{eq:background_velocity_dispersion_early_time_limit} where the velocity dispersion background simply decays in time.
\par
We generate a linear evolved cold dark matter scalar power spectrum $P_\text{CDM}(k)$ using the CLASS code \cite{blas_2011} for wave numbers $k \in \left[ 10^{-5}, \, 10 \right] h$/Mpc assuming a $\Lambda$CDM cosmology using the parameters given in table \ref{table:cosmological_parameters}. To extrapolate to smaller and larger wave numbers we use the prominent Eisenstein \& Hu fitting formula \cite{eisenstein_1999}
\begin{equation}
P_\text{CDM}(k) = \alpha \, k^{n_s} \, \frac{L^2(k)}{\big( L(k) + C(k) \, (\beta k)^2 \big)^2} \; ,
\label{eq:CDM_power_spectrum}
\end{equation}
where
\begin{equation}
L(k) = \ln(e + 1.84 \, \beta k) \; , \qquad C(k) = 14.4 + \frac{325}{1 + 60.5 \, (\beta k)^{1.11}} \; ,
\end{equation}
and $\alpha, \beta, e$ are real positive constants. 
\par
For dark matter with non-vanishing velocity dispersion, structure formation is stalled at small scales due to free-streaming of dark matter particles. In the matter dominated era, structures are erased for wave numbers larger than the comoving free-streaming wave number at matter-radiation equality $k_\text{fs,eq}$ \cite{boyanovsky_2008_1}. For wave numbers $k \gtrsim k_\text{fs,eq}$ the power spectrum is strongly suppressed and for simplicity we employ the ultraviolet cutoff $\Lambda \equiv k_\text{fs,eq}$ and use the initial scalar power spectrum
\begin{equation}
P_s^\text{in}(k) = \Theta_\text{H}^{(1)}(\Lambda - k) \, P_\text{CDM}(k) \; ,
\label{eq:initial_scalar_power_spectrum}
\end{equation}
where $\Theta_\text{H}^{(n)}(x)$ denotes the $n$-dimensional Heaviside unit step function.
\par
For our numerical investigations we consider two dark matter candidates, namely sterile neutrinos of mass $m \sim 1$ keV and weakly interacting massive particles (WIMPs) of mass $m \sim 100$ GeV. The free-streaming wave number and velocity dispersion background at matter-radiation equality are given by
\begin{equation}
k_\text{fs,eq} \approx 11 \; \frac{h}{\text{Mpc}} \; , \qquad \hat{\sigma}_\text{eq} \approx 1.1 \cdot 10^{-2} \; \frac{\text{Mpc}^2}{h^2} \; ,
\label{eq:matter-radiation_equality_values_SN}
\end{equation}
for sterile neutrino dark matter and
\begin{equation}
k_\text{fs,eq} \approx 1.6 \cdot 10^7 \; \frac{h}{\text{Mpc}} \; , \qquad \hat{\sigma}_\text{eq} \approx 5.5 \cdot 10^{-15} \; \frac{\text{Mpc}^2}{h^2} \; ,
\label{eq:matter-radiation_equality_values_WIMP}
\end{equation}
for WIMP dark matter \cite{boyanovsky_2011}. We evolve the velocity dispersion background to $\tau_\text{in}$ using the evolution equation \eqref{eq:background_velocity_dispersion_eq_v3} in an Einstein-de Sitter cosmology ($\Omega_m(\tau) = 1$, $f(\tau) = 1$) and neglect the enhancement from $\hat{Q}(\eta)$. The initial velocity dispersion background is then $\sigmahatin =( a_\text{eq} / a_\text{in}) \, \hat{\sigma}_\text{eq}$ where $a_\text{eq} \approx 1/3229$ is the scale factor at matter-radiation equality. This approximation is sensible as long as the initial time $\tau_\text{in}$ is well within the matter dominated era and $\hat{Q}(\etain) / \sigmahatin \ll 1$. Indeed, for $z(\tau_\text{in}) = 100$ we find $\hat{Q}(\etain) / \sigmahatin \approx 10^{-3}$ for sterile neutrino dark matter and $\hat{Q}(\etain) / \sigmahatin \approx 0.05$ for WIMP dark matter.

\subsection{Results}
\label{sec:results}
\begin{figure}[ht]
\centering
\begin{minipage}{.5\textwidth}
\centering
\includegraphics[scale=0.5]{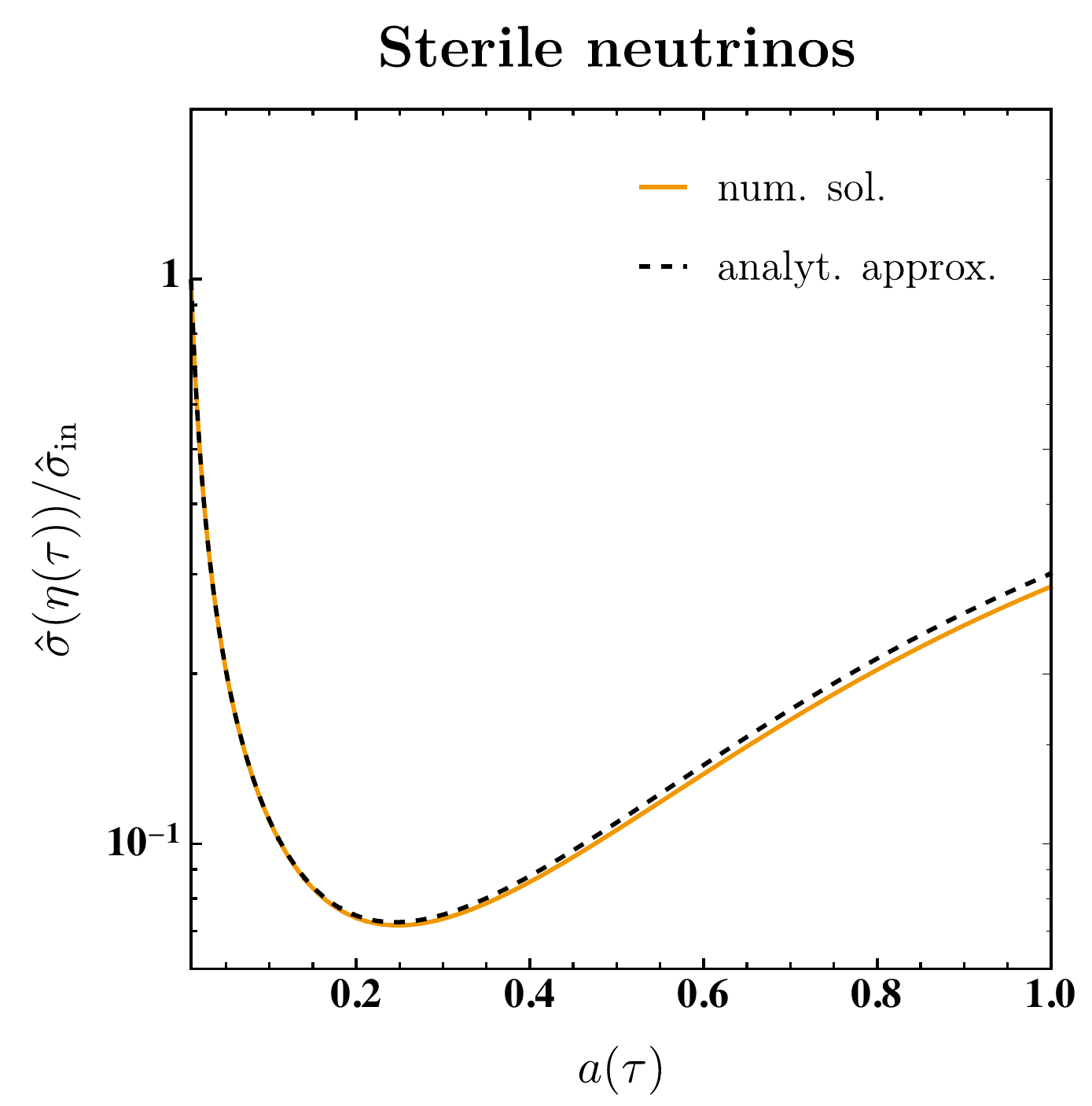}
\end{minipage}%
\begin{minipage}{.5\textwidth}
\centering
\includegraphics[scale=0.5]{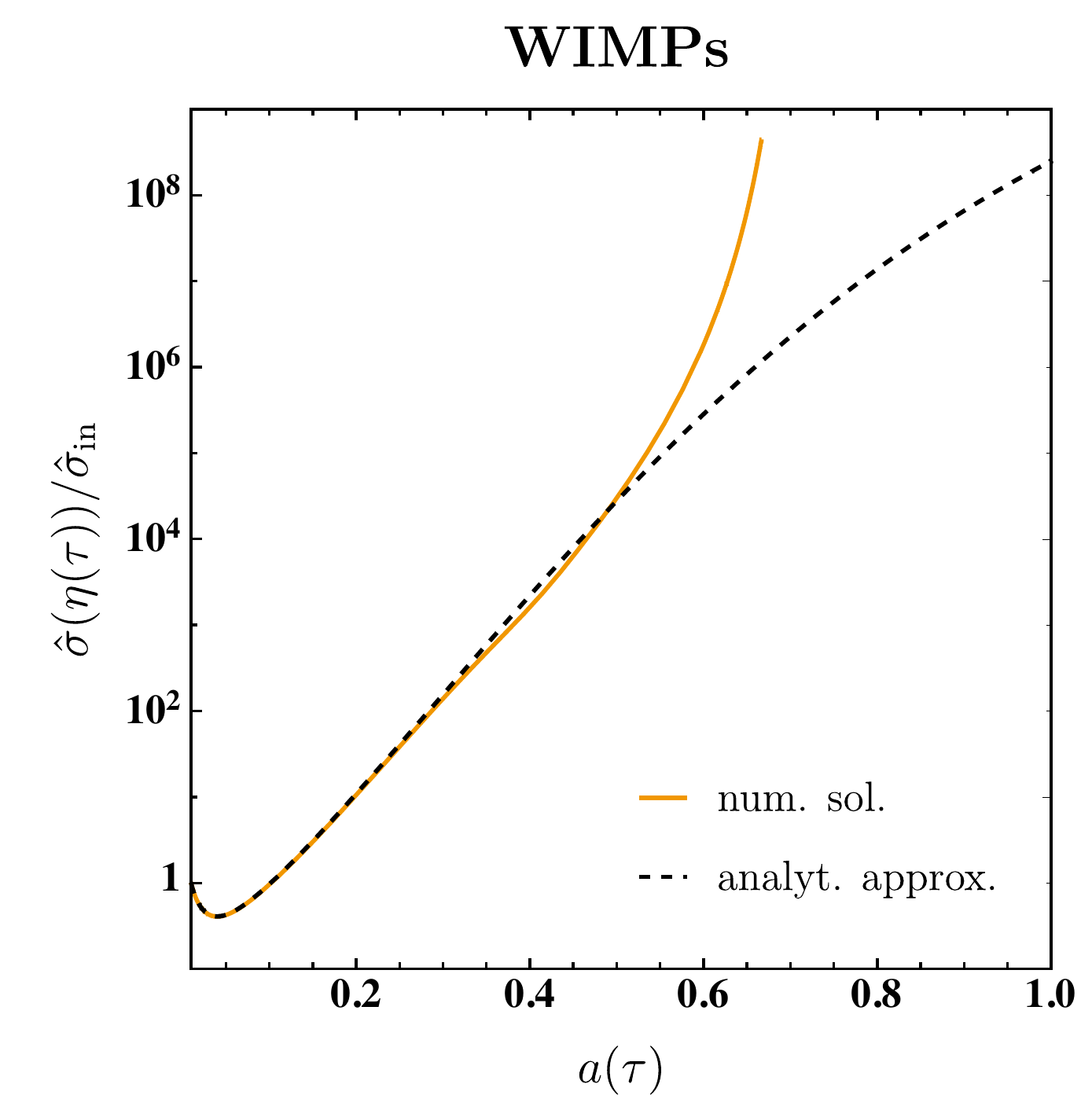}
\end{minipage}
\caption{Evolution of the velocity dispersion background $\hat{\sigma}(\eta)$ in time for sterile neutrino (left panel) and WIMP (right panel) dark matter. For sterile neutrino dark matter the numerical solution (solid yellow lines) is well approximated by the analytical approximation (dashed black lines) because the free-streaming suppression suppresses the power spectrum at small scales. For WIMP dark matter we see a strong growth and deviation from the analytical approximation at late times.}
\label{fig:background_velocity_dispersion}
\end{figure}

\begin{figure}[ht]
\centering
\begin{minipage}{.5\textwidth}
\centering
\includegraphics[scale=0.5]{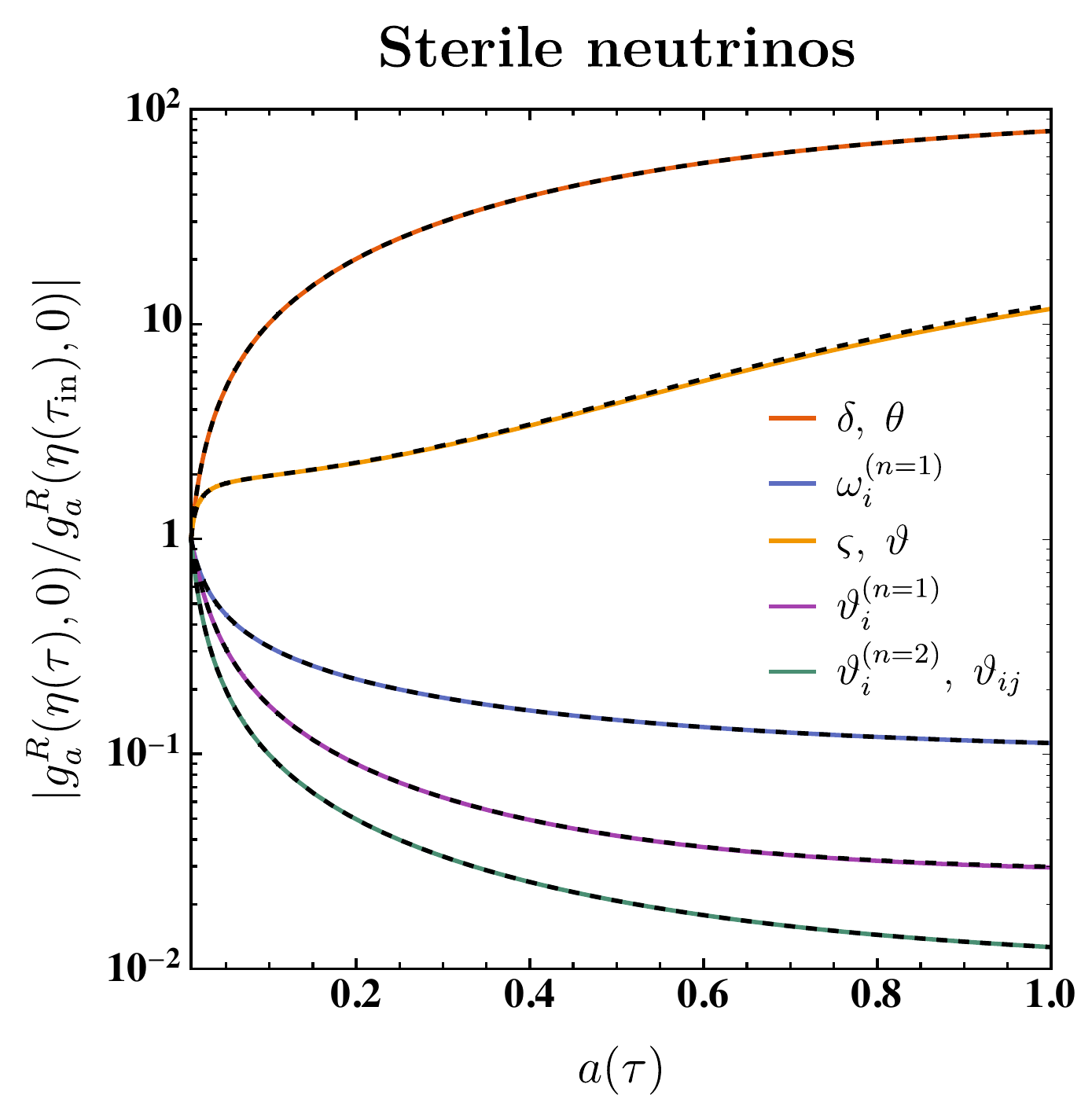}
\end{minipage}%
\begin{minipage}{.5\textwidth}
\centering
\includegraphics[scale=0.5]{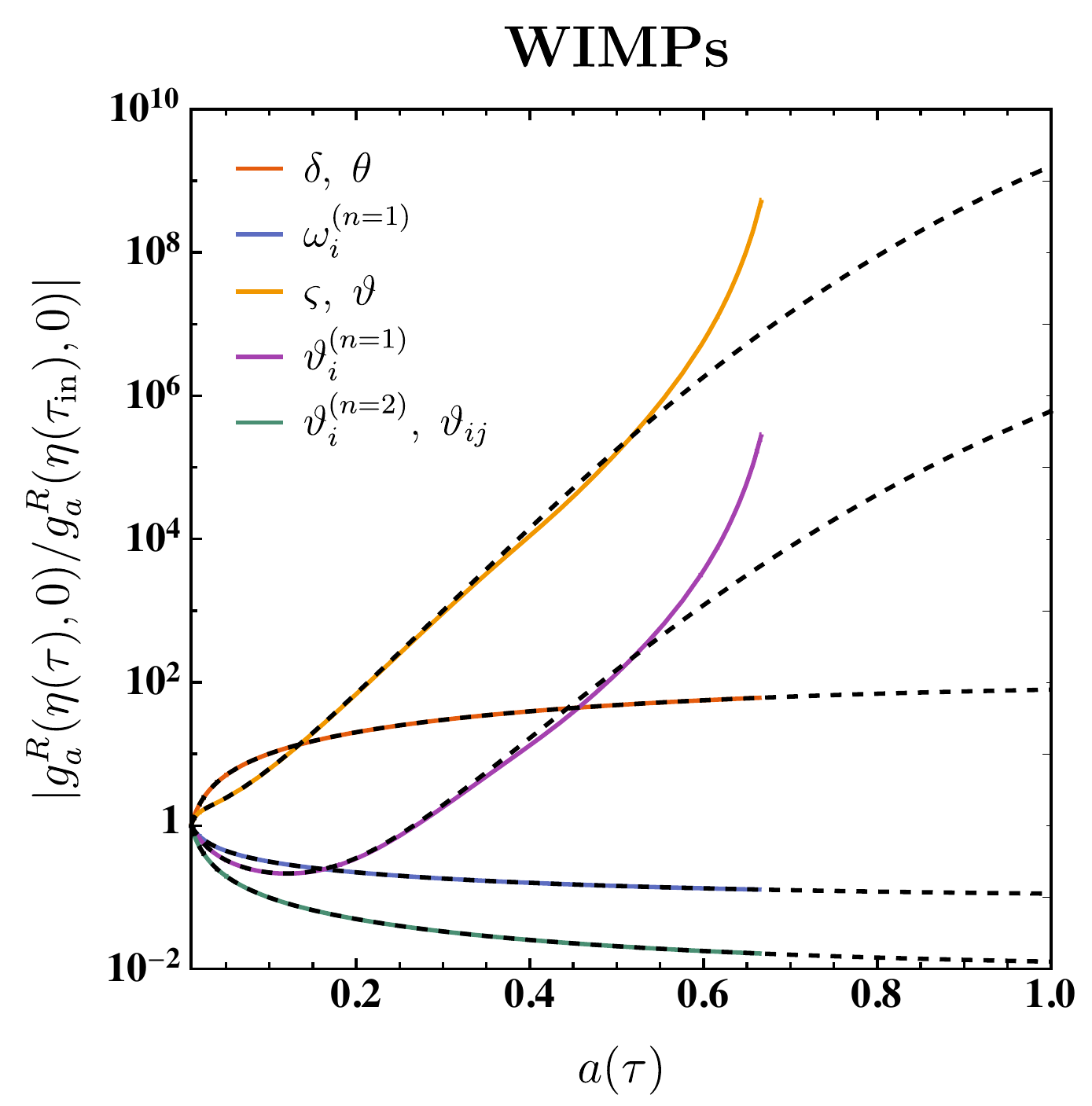}
\end{minipage}
\caption{Time evolution of the linear growth functions $\gR{a}(\eta, k)$ in the limit $k \to 0$. For sterile neutrino dark matter the scalar fluctuations obey growing solutions while the vector and tensor fluctuations decay in time. For WIMP dark matter the scalar and $n = 1$ mode vector velocity dispersion fluctuations are enhanced by the velocity dispersion background and obey a strongly growing solution at later times.}
\label{fig:linear_growth_functions_zero_mode}
\end{figure}

\begin{figure}[ht]
\centering
\begin{minipage}{.5\textwidth}
\centering
\includegraphics[scale=0.5]{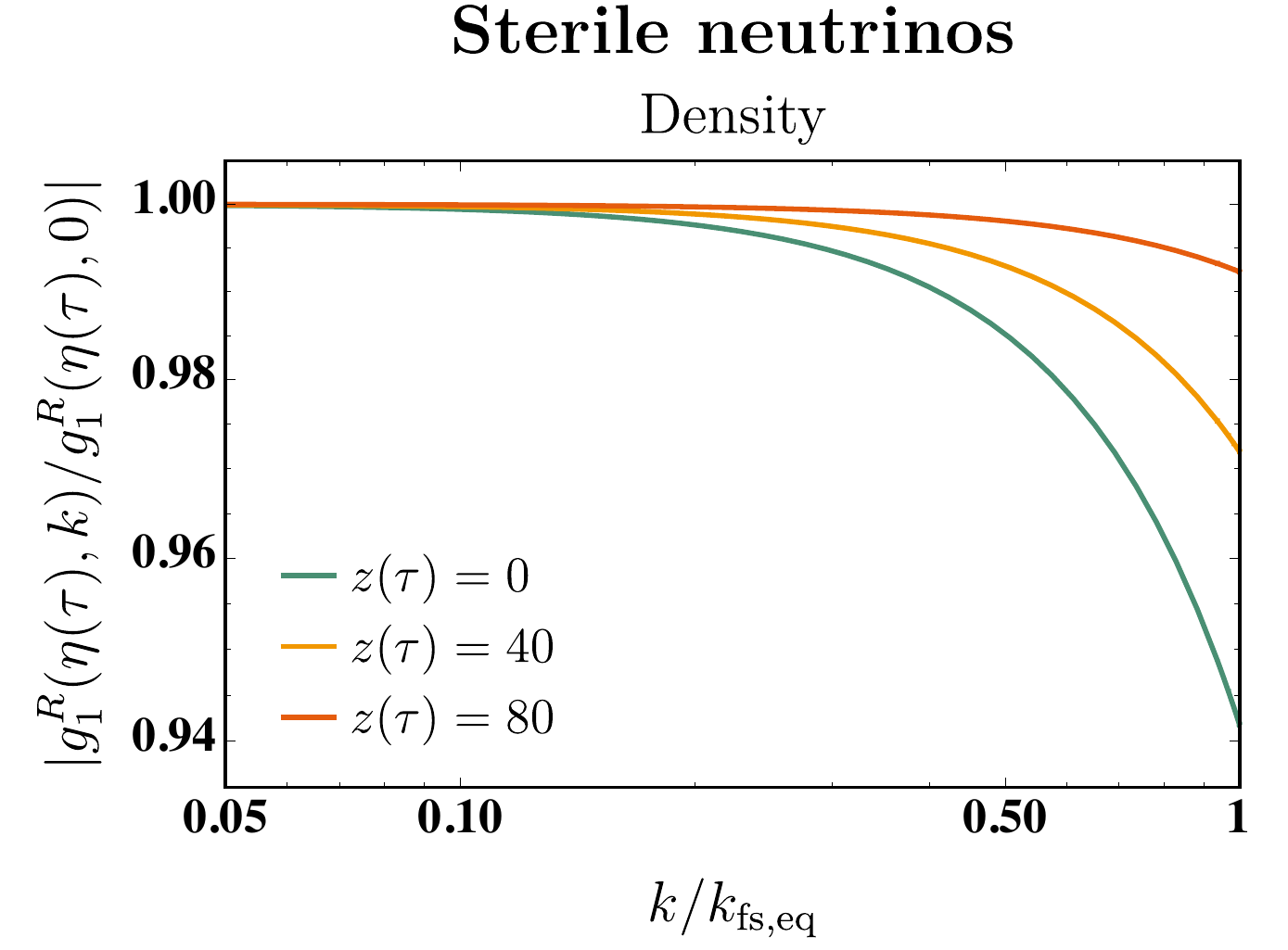}
\end{minipage}%
\begin{minipage}{.5\textwidth}
\centering
\includegraphics[scale=0.5]{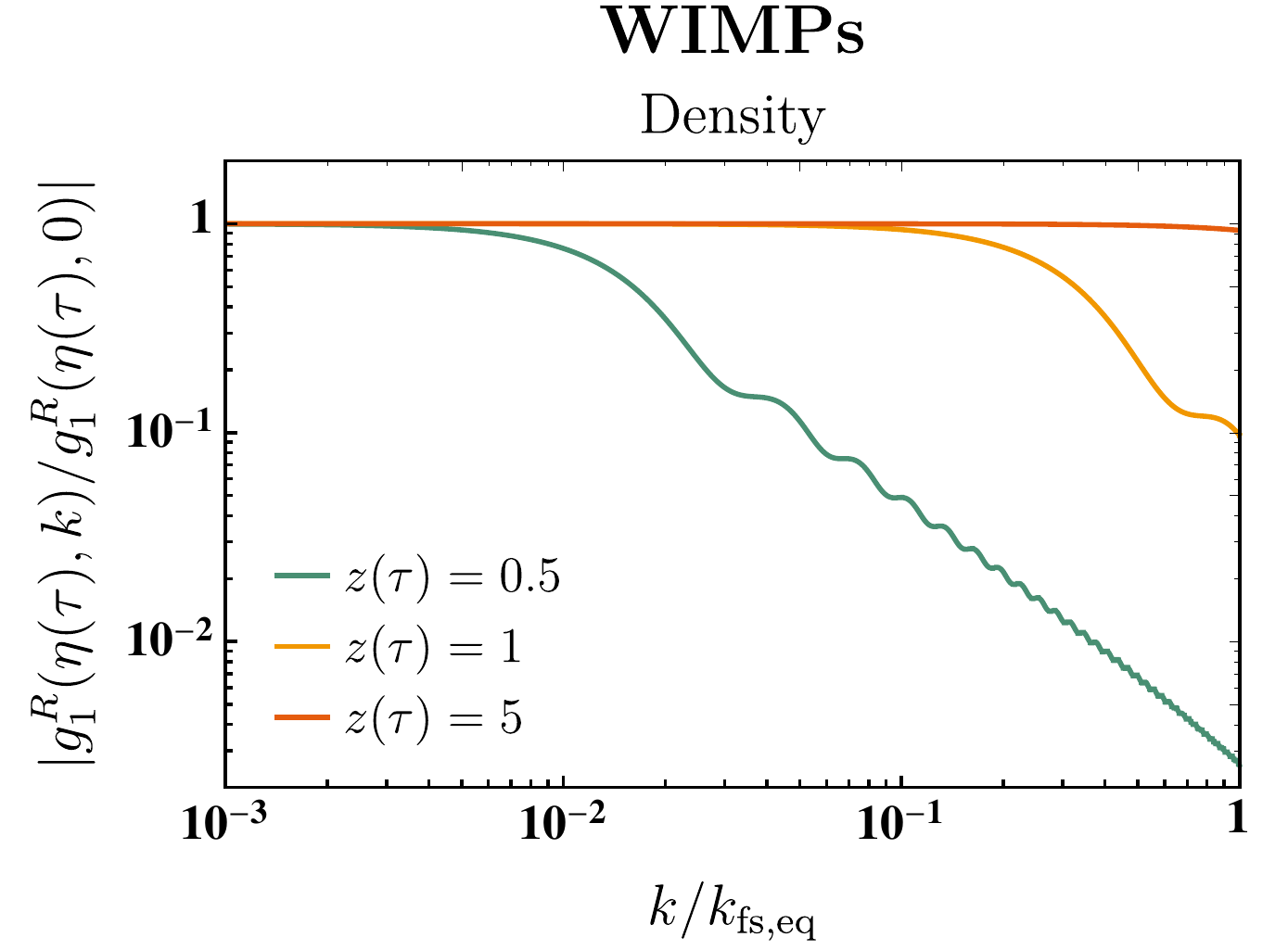}
\end{minipage}
\caption{The density linear growth function for sterile neutrino (left panel) and WIMP (right panel) dark matter normalised to the zero mode as a function of $k / \kfseq$ for three different redshifts $z(\tau)$. The density fluctuations are suppressed for larger wave numbers and lower redshift. Sterile neutrino dark matter is only weakly suppressed while WIMP dark matter shows oscillations on top of a strong suppression for wave numbers $k \le \kfseq$.}
\label{fig:density_linear_growth_function}
\end{figure}

\begin{figure}[ht]
\centering
\begin{minipage}{.5\textwidth}
\centering
\includegraphics[scale=0.5,trim={0 2cm 0 2cm},clip]{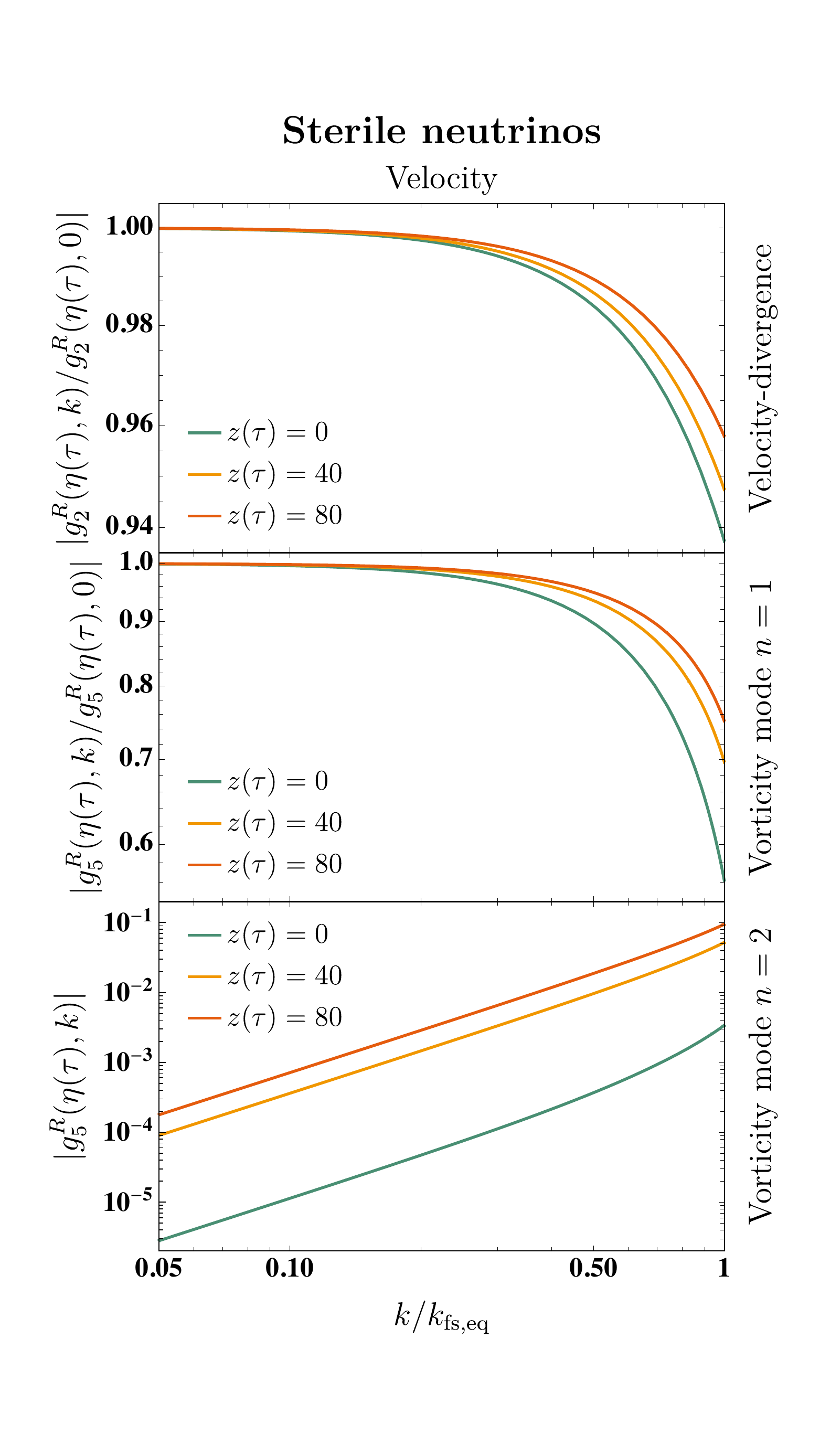}
\end{minipage}%
\begin{minipage}{.5\textwidth}
\centering
\includegraphics[scale=0.5,trim={0 2cm 0 2cm},clip]{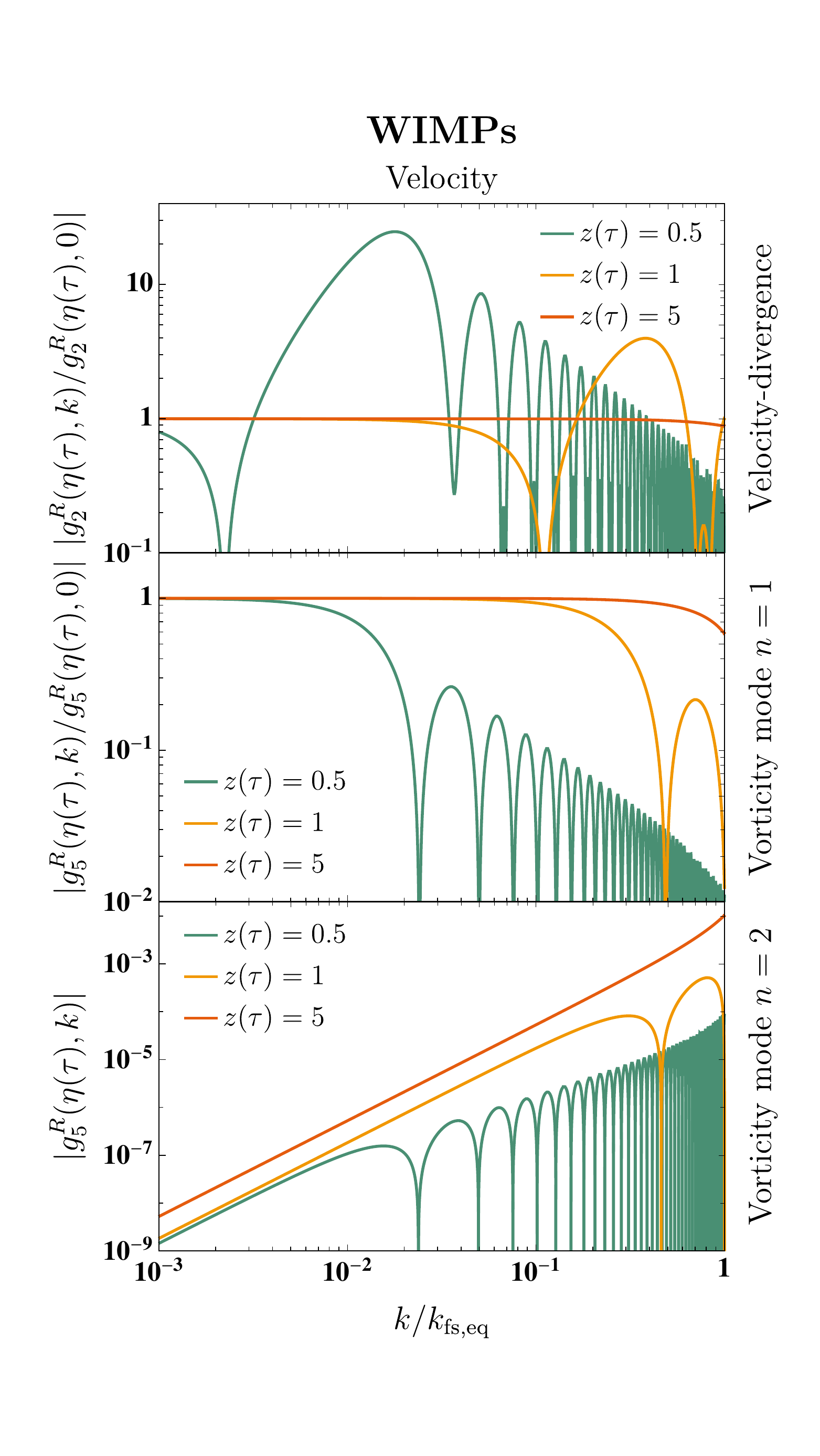}
\end{minipage}
\caption{The velocity linear growth functions for sterile neutrino (left panel) and WIMP (right panel) dark matter normalised to the zero mode as a function of $k / \kfseq$ for three different redshifts $z(\tau)$. The velocity fluctuations are suppressed for lower redshift. The velocity-divergence and $n = 1$ mode vorticity fluctuations are suppressed while the $n = 2$ mode vorticity fluctuations are enhanced for larger wave numbers. Sterile neutrino dark matter is only weakly suppressed or enhanced while WIMP dark matter shows oscillations on top of a strong suppression or enhancement for wave numbers $k \le \kfseq$.}
\label{fig:velocity_linear_growth_functions}
\end{figure}

\begin{figure}[ht]
\centering
\begin{minipage}{.5\textwidth}
\centering
\includegraphics[scale=0.5,trim={0 2cm 0 2cm},clip]{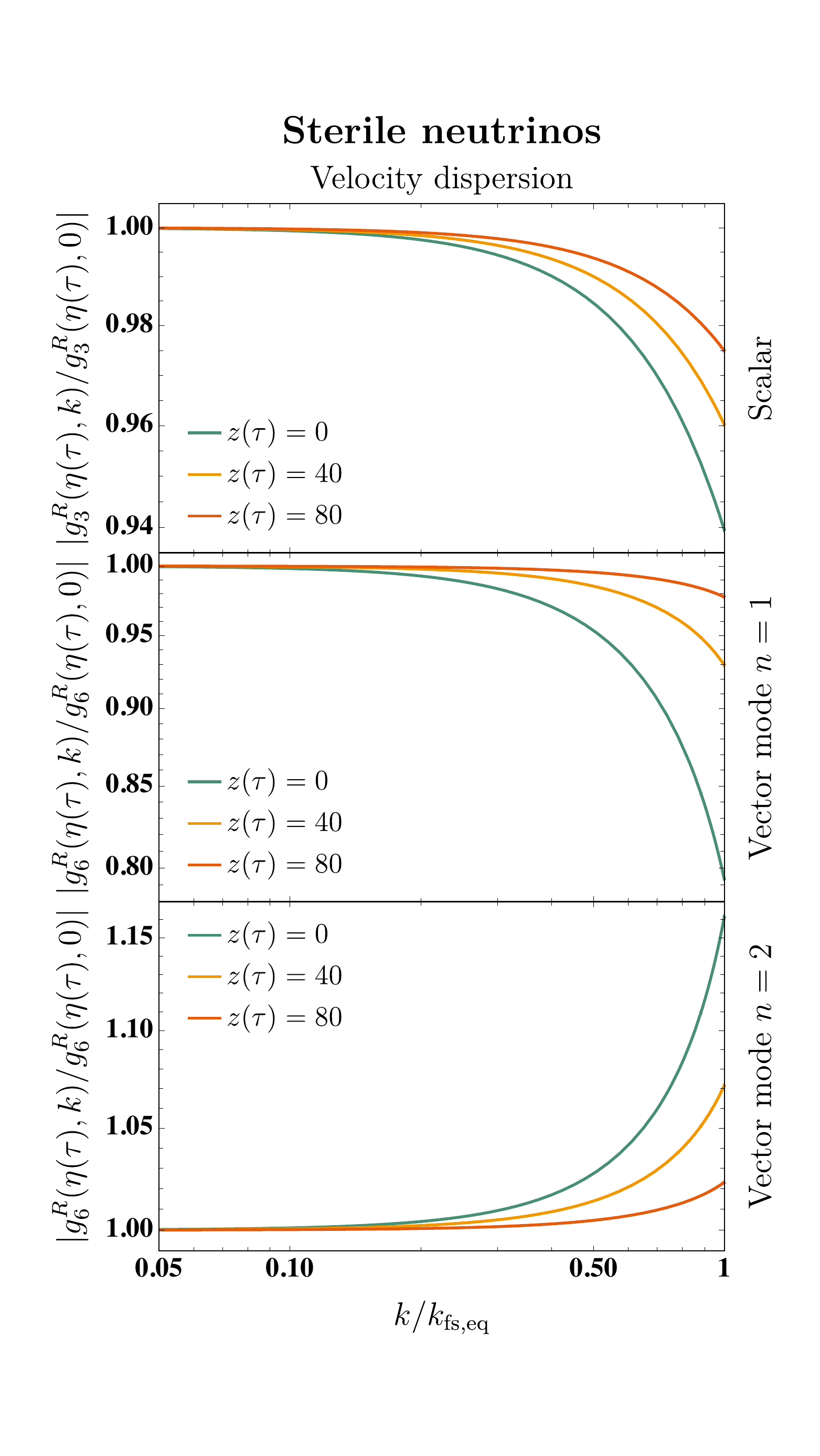}
\end{minipage}%
\begin{minipage}{.5\textwidth}
\centering
\includegraphics[scale=0.5,trim={0 2cm 0 2cm},clip]{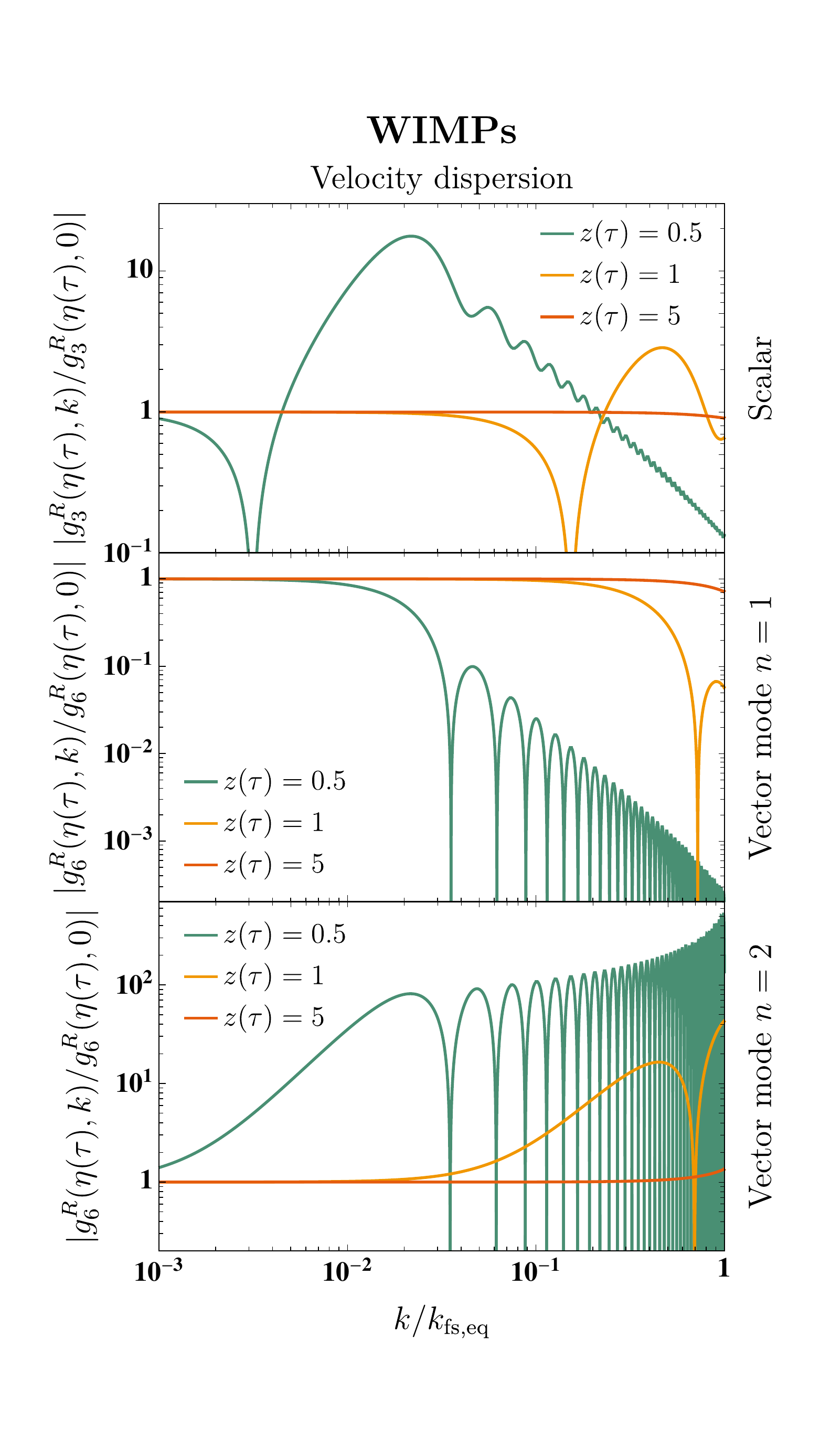}
\end{minipage}
\caption{The velocity dispersion linear growth functions for sterile neutrino (left panel) and WIMP (right panel) dark matter normalised to the zero mode as a function of $k / \kfseq$ for three different redshifts $z(\tau)$. The scalar and $n = 1$ mode vector velocity dispersion fluctuations are suppressed while the $n = 2$ mode vector velocity dispersion fluctuations are enhanced for larger wave numbers and lower redshift. Sterile neutrino dark matter is only weakly suppressed or enhanced while WIMP dark matter shows oscillations on top of a strong suppression or enhancement for wave numbers $k \le \kfseq$.}
\label{fig:velocity_dispersion_linear_growth_functions}
\end{figure}

In order to compare sterile neutrino and WIMP dark matter we show the former in the left panel and the latter in the right panel in all subsequent figures. Figures \ref{fig:background_velocity_dispersion} and \ref{fig:linear_growth_functions_zero_mode} display the time evolution of the velocity dispersion background and the zero mode of the fluctuation fields. Numerical solutions are shown as solid colored lines and the analytical approximations derived in section \ref{sec:approximation_for_small_velocity_dispersion_and_soft_power_spectrum} as dashed black lines. Figures \ref{fig:density_linear_growth_function}, \ref{fig:velocity_linear_growth_functions} and \ref{fig:velocity_dispersion_linear_growth_functions} show the scale dependence of the density, velocity and velocity dispersion fluctuation fields, respectively, at three different redshifts.
\par
Figure \ref{fig:background_velocity_dispersion} shows that initially the velocity dispersion background decays in time for both dark matter candidates, as expected for the regime \eqref{eq:background_velocity_dispersion_early_time_limit} where the contribution from fluctuations can be neglected for the evolution of $\hat{\sigma}(\eta)$. For later times however, the fluctuations grow in amplitude such that the regime \eqref{eq:background_velocity_dispersion_late_time_limit} is entered and $\hat{\sigma}(\eta)$ receives an enhancing contribution from $\hat{Q}(\eta)$. Physically speaking, the formation of inhomogeneities and anisotropies of the velocity and velocity dispersion degrees of freedom source the growth of the velocity dispersion background. From figure \ref{fig:background_velocity_dispersion} it is evident that the growth of $\hat{\sigma}(\eta)$ strongly depends on the dark matter model. For warm dark matter the free-streaming wave number $\kfseq$ is smaller than for cold dark matter and correspondingly $\hat{Q}(\eta)$ receives less contribution from larger wave numbers. For sterile neutrino dark matter the velocity dispersion background stays below its initial value while for WIMP dark matter it undergoes an immense growth at late times. For the latter we only show the evolution for $a(\tau) \le 2 / 3$ because at times where $\hat{\sigma}(\eta) \gtrsim 1 / \kfseq^2$ the validity of the analytical approximation breaks down and the source function $\hat{Q}(\eta)$ involves an integral over strongly oscillating power spectra which become difficult to control.
\par
In figure \ref{fig:linear_growth_functions_zero_mode} we show how fluctuations evolve in the large-scale limit $k \to 0$. Here the density and velocity-divergence growth functions (red) as well as the vorticity $n = 1$ mode growth function (blue) exactly obey the dynamics of the perfect pressureless fluid approximation and thus obey a growing and decaying solution, respectively.\footnote{In the limit $k \to 0$ the $n = 2$ mode of the vorticity field is not excited.} The scalar velocity dispersion growth functions (yellow) are growing in time, but depend on the evolution of the velocity dispersion background. For sterile neutrino dark matter the growth of the scalar velocity dispersion is smaller than for the density and velocity-divergence fluctuations while for WIMP dark matter the former exceeds the growth of the latter by many orders of magnitude, originating from the immense growth of $\hat{\sigma}(\eta)$. Similarly, the vector $n = 1$ mode velocity dispersion growth function (purple) is decaying in time for sterile neutrino dark matter while for WIMP dark matter the enhancement from $\hat{\sigma}(\eta)$ allows for a growing behaviour at later times. Finally, the vector $n = 1$ mode and tensor velocity dispersion growth functions (red) are independent of the velocity dispersion background and simply decay in time for both dark matter candidates.
\par
Knowing how the zero modes evolve in time we now turn to the scale dependence of the fluctuations. To this end figures \ref{fig:density_linear_growth_function}, \ref{fig:velocity_linear_growth_functions} and \ref{fig:velocity_dispersion_linear_growth_functions} display the growth functions normalised to their zero mode as a function of $k / \kfseq$. From the discussion of the growth factors in section \ref{sec:scalar_growth_factors_and_free-streaming_scale} we expect fluctuations to be suppressed at wave numbers $k \gtrsim \kfs(\eta)$ due to free-streaming. For sterile neutrino dark matter $\kfs(\eta) > \kfseq$ because $\hat{\sigma}(\eta) < \sigmahatin$ at all times and therefore the growth functions depend weakly on redshift. For WIMP dark matter on the other hand $\kfs(\eta) \ll \kfseq$ at late times because $\hat{\sigma}(\eta) \gg \sigmahatin$ and concordantly the growth functions are sensible to small changes in redshift. For sterile neutrino dark matter we show results at redshift $z(\tau) = 0$ (green line), $z(\tau) = 40$ (yellow line) and $z(\tau) = 80$ (red line) while for WIMP dark matter we choose $z(\tau) = 0.5$ (green line), $z(\tau) = 1$ (yellow line) and $z(\tau) = 5$ (red line).
\par
In figure \ref{fig:density_linear_growth_function} we show the density linear growth function. These are stronger suppressed at lower redshift and, as expected from the above discussion, for larger wave numbers. Sterile neutrino dark matter is weakly suppressed while WIMP dark matter is heavily suppressed for wave numbers $k \le \kfseq$. We observe oscillations on top of the suppression for WIMP dark matter, which are not accounted for in the growing solution growth factor nor eigenvector. As mentioned in section \ref{sec:scalar_growth_factors_and_free-streaming_scale} the eigenvectors of $\Omega_{ab}(\eta, k)$ and $\gR{ab}(\eta, k)$ are in general not identical and thus the solution dynamically evolves away from being characterised by a single growth factor. The superposition of solutions of different growth factors lead to the observed oscillations which are due to the imaginary parts of the growth factors shown in figure \ref{fig:scalar_field_growth_factors}.
\par
Figure \ref{fig:velocity_linear_growth_functions} displays the velocity-divergence (upper panel), the vorticity $n = 1$ mode (middle panel) and the vorticity $n = 2$ mode (lower panel) growth functions. Since the vorticity $n = 2$ mode growth function vanishes in the $k \to 0$ limit, we show it not normalised. The velocity-divergence and vorticity $n = 1$ mode growth functions are stronger suppressed for lower redshift and larger wave numbers. The vorticity $n = 2$ mode growth function is suppressed for lower redshift and smaller wave numbers. This behaviour qualitatively matches the behaviour of the growth factors $s_s^{(1)}(\eta, k)$ and $s_v^{(1)}(\eta, k)$ which are suppressed for larger wave numbers while $s_v^{(2)}(\eta, k)$ is enhanced. Interestingly, on top of the suppressive and oscillatory behaviour of the WIMP dark matter fluctuations discussed above we observe an enhancement of the velocity-divergence growth function in the wave number range $3 \cdot 10^{-3} \lesssim k / \kfseq \lesssim 4 \cdot 10^{-1}$. This enhancement is responsible for the strong late time growth of the velocity dispersion background and the deviation from the analytical approximation. After discussing the velocity dispersion growth functions we comment more on this, since these are needed for a complete picture of the source function $\hat{Q}(\eta)$ which depends on the velocity-velocity dispersion correlations.
\par
Figure \ref{fig:velocity_dispersion_linear_growth_functions} displays the isotropic scalar (upper panel), the vector $n = 1$ mode (middle panel) and the vector $n = 2$ mode (lower panel) velocity dispersion growth functions. The anisotropic scalar velocity dispersion growth function is not separately displayed since it is near to identical with the isotropic one. The scalar and vector $n = 1$ mode velocity dispersion growth functions are stronger suppressed for lower redshift and larger wave numbers. Contrary the vector $n = 2$ velocity dispersion mode growth function is suppressed for higher redshift and smaller wave numbers. Similar to the velocity-divergence growth function the scalar velocity dispersion growth function is enhanced in the wave number range $4 \cdot 10^{-3} \lesssim k / \kfseq \lesssim 2 \cdot 10^{-1}$. Together with the velocity-divergence growth function the strong growth of the velocity dispersion background may be qualitatively understood. For identical isotropic and anisotropic scalar velocity dispersion growth functions (at linear level these differ below the percent level) the source function \eqref{eq:Q_hat_scalar_initial_conditions} takes the form
\begin{equation}
\hat{Q}(\eta) \approx \frac{4 \pi}{3} \int_0^\infty \frac{dq \, q^2}{(2 \pi)^3} \; \gR{3}(\eta, q) \, \gR{2}(\eta, q) \, \Pin{s}(q) \; .
\end{equation}
At large wave numbers the integral runs over the heavily oscillating growth functions which are enhanced compared to the analytical approximation \eqref{eq:analytical_background_velocity_dispersion}. That is, at smaller scales the matter fluctuation fields oscillate due to the competition of gravitational collapse and free-streaming of dark matter and enhance the growth of the velocity dispersion background. Further, the suppression of the growth functions at large wave numbers is necessary for the convergence of $\hat{Q}(\eta)$. This becomes evident when taking perfectly cold initial conditions, $\Lambda \to \infty$, because the dimensionless variance \eqref{eq:definition_sigma_d} diverges in this limit. The scale dependence at linear level which induces the free-streaming suppression is therefore needed to obtain finite results.

\section{Conclusion}
In summary, we have studied an extension of the standard pressureless perfect fluid approximation for dark matter by taking the velocity dispersion tensor as an additional field into account. Neglecting all interaction of dark matter besides gravity, and assuming a kinetic description in terms of classical particles, we have obtained an equation of motion for the velocity dispersion tensor by taking the second cumulant of the Vlasov equation.
\par
The set of evolution equations is closed by neglecting the third cumulant of the local dark matter velocity distribution. In a subsequent step, we have decomposed the velocity dispersion tensor into scalar, vector and tensor fields. Together with the density and velocity fields one obtains in total four scalar fields, two solenoidal vector fields and one symmetric, transverse and traceless second-rank tensor field.
\par
Maybe the most significant modification of the single-stream approximation based on a perfect pressureless fluid picture comes from a homogeneous expectation value of the trace of the velocity dispersion tensor. On a technical level, the presence of such an expectation value has the consequence that it destroys the apparent self-consistency of the single-stream approximation. Perturbations in the scalar fields that parameterise the velocity dispersion tensor can be directly generated by other scalar perturbations. The extension to a fluid with velocity dispersion is important because it allows to account for shell-crossing and to address late times and small scales.
\par
Interestingly, the evolution equation for the velocity dispersion expectation value contains a back-reaction term at quadratic order in perturbations. Physically, this term describes how velocity dispersion grows again at late times (after a decrease due to the cosmological expansion at earlier times) due to large inhomogeneities in particular at small scales. Technically, this back-reaction effect is proportional to an integral over the power spectrum of scalar perturbations that is strongly dominated by the small scale or ultraviolet regime. Beyond our present approximation, there is also a similar contribution involving an integration over the power spectrum of vector perturbations. This ultraviolet sensitivity makes it particularly interesting to compare different dark matter candidates.
\par
We concentrate here in particular on a scenario where dark mater consists of rather light sterile neutrinos of mass $m \sim 1$ keV and weakly interacting massive particles with mass in the range $m \sim 100$ GeV. The latter are much colder than the former in the sense that they have initially, at matter-radiation equality, a much smaller velocity dispersion. However, this implies that the scalar power spectrum extends much further into the ultraviolet. As we show, the non-linear back-reaction effect is then rather strong at late times, leading to a strong increase of the velocity dispersion expectation value.
\par
A similar computation was undertaken by reference \cite{mcdonald_2011}. Here a non-vanishing velocity dispersion background was introduced and evolved in a similar way as we did, by considering one scalar velocity dispersion fluctuation field, corresponding to the combination $(\varsigma(\tau, \mathbf{k}) + \vartheta(\tau, \mathbf{k})) / 2$. Reference \cite{mcdonald_2011} found a similar suppression of the power spectrum as well as a strong late time growth of the velocity dispersion background. Further a first non-linear approximation was made by including the dominant terms which showed that the growth of the velocity dispersion background persists even for non-linear mode coupling. Another path in the Lagrangian picture was pursed by reference \cite{aviles_2016}. Here the velocity dispersion tensor was solved for perturbatively and it was found that at third order growing contributions are possible. While this approach also features a velocity dispersion background as the zeroth order solution, it differs from our approach in the sense that we solve the background evolution equation fully non-linear, while solving the fluctuation fields linearly. The feature of a suppressed power spectrum due to a characteristic free-streaming scale was also found by reference \cite{aviles_2016}.
\par
In the present paper we have concentrated on a relatively simple approximation where the background is evolved non-linearly including the back-reaction effects from integrals over the power spectra of perturbations. The perturbations, or deviations from the homogeneous and isotropic background, are propagated linearly, however. While many qualitative effects can be well studied on this basis, one should keep in mind that non-linear effects in the evolution of perturbations can change the picture quantitatively, in particular at small scales and late times. Because of the ultraviolet dominance described above, this can also affect the evolution of the velocity dispersion expectation value. In future work we plan to take such effects into account. This could be done perturbatively but also using resummation schemes \cite{valageas_2004, crocce_2006_1, crocce_2006_2, mcdonald_2007, valageas_2007, matarrese_2007, izumi_2007, bernardeau_2008_2, taruya_2008, crocce_2008, matsubara_2008_1, valageas_2008, bernardeau_2008_1, matsubara_2008_2, pietroni_2008, taruya_2009, bernardeau_2010, anselmi_2011, bernardeau_2012_1, bernardeau_2012_2, taruya_2012, anselmi_2012, bernardeau_2013} and the non-perturbative renormalisation group \cite{matarrese_2008, floerchinger_2017}. We are particularly motivated by the perspective that such a theoretical framework can go beyond the limitations of the conventional pressureless perfect fluid approximation when it comes to shell-crossing at late times and small scales. This could extend the range of applicability of semi-analytical techniques for cosmological structure formation substantially and thereby lead to an improved understanding of dark matter and late time cosmology.

\appendix
\section{Non-linear terms of the equations of motion}
\label{non-linear_terms_of_the_equations_of_motion}
The terms quadratic in the fluctuation fields in the continuity equation \eqref{eq:continuity_eq_v2} are abbreviated by
\begin{equation}
I_\delta(\tau, \mathbf{k}) \equiv \int_{\mathbf{k}_1, \mathbf{k}_2} \Big[ A_{\theta \delta} \, \theta_1 \delta_2 + A_{\omega \delta}^i \, {\omega_i}_1 \delta_2 \Big] \; .
\label{eq:quadratic_terms_continuity_eq}
\end{equation}
Similarly, for the velocity-divergence and vorticity equations \eqref{eq:velocity-divergence_equation} and \eqref{eq:vorticity_equation} we abbreviate
\begin{equation}
\begin{aligned}
I_\theta(\tau, \mathbf{k}) \equiv& \int_{\mathbf{k}_1, \mathbf{k}_2} \bigg[ B_{\theta \theta} \, \theta_1 \theta_2 + B_{\theta \omega}^i \, \theta_1 {\omega_{i}}_2 + B_{\omega \omega}^{ij} \, {\omega_i}_1 {\omega_j}_2 + k^2 \bar{\sigma} \, B_{\delta \delta} \, \delta_1 \delta_2 \\
&\hspace*{1.4cm} + B_{\varsigma \delta} \, \varsigma_1 \delta_2 + B_{\vartheta \delta} \, \vartheta_1 \delta_2 + B_{\vartheta \delta}^i \, {\vartheta_i}_1 \delta_2 + B_{\vartheta \delta}^{ij} \, {\vartheta_{ij}}_1 \delta_2 \bigg] \; , \\
I_\omega^i(\tau, \mathbf{k}) \equiv& \int_{\mathbf{k}_1, \mathbf{k}_2} \bigg[ C_{\theta \omega}^{ij} \, \theta_1 {\omega_j}_2 + C_{\omega \omega}^{ijk} \, {\omega_j}_1 {\omega_k}_2 + C_{\varsigma \delta}^i \, \varsigma_1 \delta_2 \\
&\hspace*{1.4cm} + C_{\vartheta \delta}^i \, \vartheta_1 \delta_2 + C_{\vartheta \delta}^{ij} \, {\vartheta_j}_1 \delta_2 + C_{\vartheta \delta}^{ijk} \, {\vartheta_{jk}}_1 \delta_2 \bigg] \; ,
\end{aligned}
\end{equation}
and in the velocity dispersion equations \eqref{eq:trace_velocity_dispersion_eq} -- \eqref{eq:tensor_velocity_dispersion_eq} we use
\begin{equation}
\begin{aligned}
I_\varsigma(\tau, \mathbf{k}) \equiv& \int_{\mathbf{k}_1, \mathbf{k}_2} \Big[ D_{\varsigma \theta} \, \varsigma_1 \theta_2 + D_{\varsigma \omega}^i \, \varsigma_1 \omega_{i \, 2} + D_{\vartheta \theta} \, \vartheta_1 \theta_2 + D_{\vartheta \omega}^i \, \vartheta_1 \omega_{i \, 2} \\
& \hspace*{1.3cm} + D_{\vartheta \theta}^i \, \vartheta_{i \, 1} \theta_2 + D_{\vartheta \omega}^{ij} \, \vartheta_{i \, 1} \omega_{j \, 2} + D_{\vartheta \theta}^{ij} \, \vartheta_{ij \, 1} \theta_2 + D_{\vartheta \omega}^{ijk} \, \vartheta_{ij \, 1} \omega_{k \, 2} \Big] \; , \\
I_{\vartheta}(\tau, \mathbf{k}) \equiv& \int_{\mathbf{k}_1, \mathbf{k}_2} \Big[ E_{\varsigma \theta} \, \varsigma_1 \theta_2 + E_{\varsigma \omega}^i \, \varsigma_1 \omega_{i \, 2} + E_{\vartheta \theta} \, \vartheta_1 \theta_2 + E_{\vartheta \omega}^i \, \vartheta_1 \omega_{i \, 2} \\
& \hspace*{1.3cm} + E_{\vartheta \theta}^i \, \vartheta_{i \, 1} \theta_2 + E_{\vartheta \omega}^{ij} \, \vartheta_{i \, 1} \omega_{j \, 2} + E_{\vartheta \theta}^{ij} \, \vartheta_{ij \, 1} \theta_2 + E_{\vartheta \omega}^{ijk} \, \vartheta_{ij \, 1} \omega_{k \, 2} \Big] \; , \\
I_{\vartheta}^i(\tau, \mathbf{k}) \equiv& \int_{\mathbf{k}_1, \mathbf{k}_2} \Big[ F_{\varsigma \theta}^i \, \varsigma_1 \theta_2 + F_{\varsigma \omega}^{ij} \, \varsigma_1 \omega_{j \, 2} + F_{\vartheta \theta}^i \, \vartheta_1 \theta_2 + F_{\vartheta \omega}^{ij} \, \vartheta_1 \omega_{j \, 2} \\
& \hspace*{1.3cm} + F_{\vartheta \theta}^{ij} \, \vartheta_{j \, 1} \theta_2 + F_{\vartheta \omega}^{ijk} \, \vartheta_{j \, 1} \omega_{k \, 2} + F_{\vartheta \theta}^{ijk} \, \vartheta_{jk \, 1} \theta_2 + F_{\vartheta \omega}^{ijkl} \, \vartheta_{jk \, 1} \omega_{l \, 2} \Big] \; , \\
I_{\vartheta}^{ij}(\tau, \mathbf{k}) \equiv& \int_{\mathbf{k}_1, \mathbf{k}_2} \Big[ G_{\varsigma \theta}^{ij} \, \varsigma_1 \theta_2 + G_{\varsigma \omega}^{ijk} \, \varsigma_1 \omega_{k \, 2} + G_{\vartheta \theta}^{ij} \, \vartheta_1 \theta_2 + G_{\vartheta \omega}^{ijk} \, \vartheta_1 \omega_{k \, 2} \\
& \hspace*{1.3cm} + G_{\vartheta \theta}^{ijk} \, \vartheta_{k \, 1} \theta_2 + G_{\vartheta \omega}^{ijkl} \, \vartheta_{k \, 1} \omega_{l \, 2} + G_{\vartheta \theta}^{ijkl} \, \vartheta_{kl \, 1} \theta_2 + G_{\vartheta \omega}^{ijklm} \, \vartheta_{kl \, 1} \omega_{m \, 2} \Big] \; .
\end{aligned}
\end{equation}
Finally, the terms cubic in the fluctuation fields are abbreviated as
\begin{equation}
\begin{aligned}
J_\theta(\tau, \mathbf{k}) \equiv& \int_{\mathbf{k}_1, \mathbf{k}_2, \mathbf{k}_3} \Big[ k^2 \bar{\sigma} \, B_{\delta \delta \delta} \, \delta_1 \delta_2 \delta_3 + B_{\varsigma \delta \delta} \, \varsigma_1 \delta_2 \delta_3 \\
&\hspace*{2.4cm} + B_{\vartheta \delta \delta} \, \vartheta_1 \delta_2 \delta_3 + B_{\vartheta \delta \delta}^i \, {\vartheta_i}_1 \delta_2 \delta_3 + B_{\vartheta \delta \delta}^{ij} \, {\vartheta_{ij}}_1 \delta_2 \delta_3 \Big] \; , \\
J_\omega^i(\tau, \mathbf{k}) \equiv& \int_{\mathbf{k}_1, \mathbf{k}_2, \mathbf{k}_3} \left[ C_{\varsigma \delta \delta}^i \, \varsigma_1 \delta_2 \delta_3 + C_{\vartheta \delta \delta}^i \, \vartheta_1 \delta_2 \delta_3 + C_{\vartheta \delta \delta}^{ij} \, {\vartheta_j}_1 \delta_2 \delta_3 + C_{\vartheta \delta \delta}^{ijk} \, {\vartheta_{jk}}_1 \delta_2 \delta_3 \right] \; .
\end{aligned}
\label{eq:cubic_terms_vorticity_eq}
\end{equation}
All fields are evaluated at time $\tau$ and in order to simplify notation we denote a field with wave vector argument $\mathbf{k}_1$, $\mathbf{k}_2$ or $\mathbf{k}_3$ by a subscripted $1$, $2$ or $3$, respectively. The vertices $A, B, C, D, E, F$ and $G$ are the convolution kernels of the Fourier transformation of the non-linear terms of the equations of motion of the corresponding fluctuation fields. They carry an overall momentum conserving Dirac delta function and depend on the wave vectors which are integrated over. We provide the explicit form of these vertices in a future publication.

\acknowledgments{This work is part of the DFG Collaborative Research Centre ``SFB 1225 (ISOQUANT)'' as well as the DFG project BE 2795/4-1 and EXC-2181/1 -- 390900948 (the Heidelberg
STRUCTURES Excellence Cluster).}

\end{document}